\documentclass[prx, twocolumn, groupedaddress]{revtex4-1}
\usepackage{graphicx,color}
\usepackage{amssymb}   
\usepackage{amsmath}
\usepackage{amsfonts}
\usepackage{mathdots}
\usepackage{hyperref}
\definecolor{myblue}{RGB}{46, 48,146}
\hypersetup{colorlinks=true,linkcolor=myblue,citecolor=myblue,urlcolor=myblue, linktocpage}
\usepackage{braket}
\usepackage{bm}
\usepackage{natbib}
\usepackage{gensymb}
\usepackage{threeparttable}
\usepackage{multirow}
\usepackage{longtable}
\usepackage{booktabs}
\usepackage{gensymb}

\newcommand{\vect}[1]{\boldsymbol{#1}}

\begin{document}
\title{Equivalence of semiclassical and response theories for second-order nonlinear ac Hall effects}
\author{Jinxiong Jia$^{1,2}$}
\thanks{These authors contributed equally to this work.}
\author{Longjun Xiang$^1$}
\thanks{These authors contributed equally to this work.}
\author{Zhenhua Qiao$^2$}
\author{Jian Wang$^{1,2,3}$}
\email[]{jianwang@hku.hk}
\affiliation{College of Physics and Optoelectronic Engineering, Shenzhen University, Shenzhen 518060, China}
\affiliation{Department of Physics, University of Science and Technology of China, Hefei, Anhui 230026, China}
\affiliation{Department of Physics, The University of Hong Kong, Pokfulam Road, Hong Kong, China}

\begin{abstract}
It has been known that the semiclassical theory and the response theory can
equivalently give the Drude and the intrinsic anomalous Hall conductivities in the linear order of electric field.
However, recent theoretical advances implied that
the second-order nonlinear conductivities calculated with both approaches
are no longer equivalent, which leads to various experimental explanations even in a similar experimental setup
conducted in 
\href{https://www.science.org/doi/10.1126/science.adf1506}{[\textit{Science \textbf{381}, 181 (2023)}]}
and 
\href{https://www.nature.com/articles/s41586-023-06363-3}{[\textit{Nature \textbf{621}, 487 (2023)}]}, respectively.
Herein, by extending the AC semiclassical theory up to the second order of electric field,
we show that the semiclassical theory is still equivalent to the response theory in the second order of electric field
when the relaxation is taken into account on the same footing.
In particular, we show that the familiar second-order nonlinear current responses,
including the nonlinear Drude current and
the Berry curvature (quantum metric) dipole driven extrinsic (intrinsic) nonlinear Hall current,
can be derived by both approaches.
Further, we show that the quantum-corrected intrinsic nonlinear longitudinal current,
as recently proposed by the response theory or in a similar manner,
can also be reproduced by the semiclassical theory.
Beyond those known second-order current responses,
with both approaches, we uncover two previously overlooked
nonlinear displacement currents unique to the AC electric field.
As a consequence of this equivalence,
(i) we suggest that
the energy of the equilibrium Fermi distribution particularly in the semiclassical theory
should be the unperturbed one by assuming that both approaches give the same intrinsic responses;
(ii) we unify the intrinsic second-order nonlinear longitudinal
current responses calculated with both approaches;
(iii) we argue that the scheme of introducing relaxation in the response theory
by modifying the quantum Liouville equation needs to be reconsidered.
Our work explicitly shows the equivalence of the AC semiclassical theory and the response theory
when calculating the second-order nonlinear conductivities under the electric field and highlights 
the influence of the AC electric field.
\end{abstract}

\maketitle


\section{Introduction}
Understanding the response behaviors of Bloch electrons in crystalline solids
under the electromagnetic field is one of the oldest but still most active themes
in condensed matter physics,
which may be traced back to the phenomenological Drude model for metal with a finite Fermi surface \cite{Mermin}.
Along this way, by treating the Bloch electron as a wavepacket which
satisfies the time-dependent Schr\"odinger equation,
the semiclassical equations of motion under the static electric field $\vect{E}$
\cite{Niu1996, Niu1999, Xiao2010, Bfield}
($e=\hbar=1$)
\begin{align}
\dot{\vect{r}} &= - \dot{\vect{k}} \times \vect{\Omega}_n + \partial_{\vect{k}} \epsilon_n,
\label{velocity}
\\
\dot{\vect{k}} &= - \vect{E} ,
\label{momentum}
\end{align}
are proposed to evaluate the velocity of the wavepacket in phase space.
In Eq.~(\ref{velocity}), $\dot{\vect{r}}$ and $\dot{\vect{k}}$
denotes the time derivative of the position and momentum of the wavepacket, respectively.
In addition, $\vect{\Omega}_n$ and $\epsilon_n$ stand for the Berry curvature and band energy
for the $n$th band, respectively.
Further, by combining with the Boltzmann equation for the nonequilibrium distribution function $\bar{f}_n$
for the $n$th band,
\begin{align}
\vect{E} \cdot \partial_{\vect{k}} \bar{f}_{n} - \partial_t \bar{f}_n = (\bar{f}_n-f_n)/\tau,
\label{boltz0}
\end{align}
where $f_n$ is the equilibrium Fermi distribution function,
the current at the linear and nonlinear regimes can be calculated
\cite{BCD, GaoY2014, GaoY2019, BPT, KTLawthird, KTLaw, XiaoCnc, E0Ew1, E0Ew2},
where the scattering is usually introduced by adopting the relaxation time ($\tau$) approximation.
Importantly, the linear and nonlinear conductivity tensors usually are expressed by
the quantum Bloch wavefunction and thereby the detected current
can manifest the quantum geometry
\cite{QuantumGeometry, YanPRR, QuantumGeometry1, QuantumGeometry2, MaQlight, JTsong, AMIT}
and topological properties of quantum materials
\cite{topology, Nagaosa2017, JEMoore2017}.

Eqs.(\ref{velocity}-\ref{boltz0}), especially when combining with the well-developed
first-principles material calculations \cite{Yanfirst},
offers one of the most standard and successful theoretical tools to quantitatively 
understand the exotic linear and nonlinear transport experiments,
such as the intrinsic anomalous Hall effect \cite{NiuQ2002, FZ2003, YaoYG2004, Nagaosa2010}
in ferromagnetic metals and the extrinsic nonlinear Hall effect
in nonmagnetic quantum materials \cite{BCDexp1, BCDexp2, BCDexp3},
where the Berry curvature $\vect{\Omega}_n$ plays a pivotal role \cite{Xiao2010}.
Recently, based on the second-order DC semiclassical theory \cite{GaoY2014},
one further realizes that the counterpart of the Berry curvature in the quantum geometric tensor---quantum metric,
can also play a critical role, such as in 
the intrinsic second-order nonlinear Hall effect
\cite{intrinsic20211, intrinsic20212, XuSY2023, Wang2023, MNHE1, MNHE2, QMNP}
and the extrinsic third-order nonlinear Hall effect \cite{BPT, KTLawthird}.
These progresses spark the quantum geometric physics \cite{QuantumGeometry} in topological quantum materials.

Compared to the transport in metal, the responses of Bloch electrons in insulators with
a vanishing Fermi surface are very often explored by irradiating light (an AC electric field),
which may be traced back to the phenomenological Lorentz model \cite{Lorentz}
for the dielectric solids. Up to date, the more standard and mature approach
used to tackle the same problem is the response theory
(typically also referred to as the quantum kinetic theory)
\cite{Kraut1979,Baltz1981,Belinicher1980,Sipe1995,Sipe2000,Rappe2012,JEMoore2019,Yan2019,Wanghua2020,Rappe2023, Juan2020,
WatanabePRX,Bequ}
where the valence electrons by absorbing a photon to form a \textit{resonant} conducting current can be well
captured. Within the framework of the response theory,
the linear and nonlinear currents are
obtained by solving the (quantum) Liouville equation
for the nonequilibrium density matrix $\rho$
\cite{Sipe1995, Sipe2000, JEMoore2019},
\begin{align}
i \partial_t \rho = [H, \rho],
\label{QL0}
\end{align}
where $H=H_0+H_1$ with $H_0$ being the Hamiltonian of the periodic solid
and
$H_1=\vect{E}(t)\cdot\vect{r}$ the perturbed electric field at the length gauge \cite{Sipe1995, Sipe2000, vgauge}.

Similarly, by combining with the well-developed first-principles material calculations,
the expressions derived from the response theory can also be used
to quantitatively evaluate the current in realistic materials \cite{Rappe2012, Yan2019, Wanghua2020, Rappe2023},
such as the shift and injection ones \cite{Sipe1995, Sipe2000, Rappe2012, Yan2019, Wanghua2020, Rappe2023}.
Notably, a large number of experiments under light illumination
can be explained by those first-principles calculations
\cite{relax, Hualingzeng, jerk1, jerk2, shiftexp1, circularexp1}.
In addition, these exotic responses can also be employed to diagnose the
quantum geometric properties 
\cite{QuantumGeometry, YanPRR, QuantumGeometry1, QuantumGeometry2, MaQlight, JTsong, AMIT}
of Bloch electrons in quantum materials.

Despite their distinct starting points, the semiclassical theory and the response theory
has been complementary in many aspects
\cite{Luttinger1958, Sinitsyn2007PRB, Sinitsyn2007, XiangDHE,
Yanase2020, Yanunify, YBH2ndenergy, CulcerPRBL, GangSu2022, XiaoC2019PRB, intrinsicThird, Sodemann2019}.
For instance, both formulations give the same linear response coefficients
\cite{Luttinger1958, Sinitsyn2007PRB, Sinitsyn2007, XiangDHE}.
In addition, all the nonlinear (charge) response relations derived from the semiclassical theory,
which contains the extrinsic nonlinear Drude current,
the extrinsic nonlinear Hall current \cite{BCD},
and the intrinsic nonlinear Hall current \cite{GaoY2014, intrinsic20211, intrinsic20212},
have been reproduced by the response theory under the DC limit \cite{Yanase2020, Sodemann2019}.
However, an intrinsic second-order nonlinear longitudinal current
(nonreciprocal magnetoresistance),
which is believed to be beyond the semiclassical results
given in the pioneering semiclassical paper Ref. \cite{GaoY2014},
is proposed \cite{YBH2ndenergy, CulcerPRBL, GangSu2022} based on the response theory
or a similar method,
which together with the intrinsic nonlinear Hall effect
has been observed experimentally in Ref. \cite{Wang2023}.
Interestingly, in a similar experimental setup,
the intrinsic nonlinear Hall effect is also observed in Ref. \cite{XuSY2023},
where the nonreciprocal magnetoresistance does not appear,
as consistent with Ref. \cite{GaoY2014}.
Motivated by these progresses, an exhausted comparison between the semiclassical theory
and the response theory becomes urgent.

To that end, two remarks are given in order.
First, we note that Eqs. (\ref{velocity}-\ref{momentum})
under the static electric field have been directly extended to calculate
the time-dependent high-frequency responses
\cite{BCD, Terahertz0, Terahertz1, Terahertz2}.
Meanwhile, almost all the experiments
\cite{BCDexp1, BCDexp2, BCDexp3, XuSY2023, Wang2023, MNHE1}
to verify these semiclassical proposals are performed by applying an AC driving field.
However, the complete semiclassical equations of motion
under the AC (time-dependent) electric field read \cite{Niu1996, Niu1999, Xiao2010}
\begin{align}
\dot{\vect{r}} &= - \dot{\vect{k}} \times \vect{\Omega}_n + \partial_{\vect{k}} \epsilon_n - \vect{\Omega}_n^{\vect{k}t},
\label{velocityAC}
\\
\dot{\vect{k}} &= - \vect{E} (t) ,
\label{momentumAC}
\end{align}
where the mixed Berry curvature $\vect{\Omega}_n^{\vect{k}t}$ defined in momentum-time space,
like the conventional Berry curvature $\vect{\Omega}_n$,
also shows a critical role,
such as in the adiabatic charge pumping \cite{chargepump} and the emergent electromagnetic induction \cite{NagaosaAC}.
It has been clear that the electric field can lead to
a correction to the conventional Berry curvature $\vect{\Omega}_n$ and
further induces the intrinsic nonlinear Hall effect
\cite{GaoY2014, intrinsic20211, intrinsic20212, XuSY2023, Wang2023},
but how the mixed Berry curvature can be modified by the electric field
and whether or not this modification can bring up any new physics remains unknown.
Furthermore, an explicit comparison between the semiclassical theory 
and the response theory with a finite frequency can not be conducted,
particularly in the second order of the electric field.

Second, the inclusion of relaxation in the response theory usually was achieved
by phenomenologically modifying the Liouville equation as \cite{Mikhailov, Peres}
\begin{align}
i\partial_t \rho = [H, \rho] - \frac{i}{\tau} (\rho - \rho^{(0)}),
\label{res1}
\end{align}
where $\rho^{(0)}$ is the equilibrium density matrix.
Although Eq. (\ref{res1}) has been used in many studies
to discuss the nonlinear responses,
the last term in Eq. (\ref{res1}) has only been partially understood
\cite{Sodemann2019, HWXu2021, Culcer2022, disorderCulcer1,disorderCulcer2,disorderCulcer3,Agawarl},
and an explicit comparison with the relaxation mechanism of the semiclassical theory remains elusive.

To address these challenges, we extend the semiclassical theory under an AC uniform electric field
up to the second order of the electric field.
We show that up to the second order of electric field,
the AC semiclassical theory is equivalent to the response theory in the clean limit.
Remarkably, this equivalence can be inherited when the relaxation,
which is introduced by solving the Boltzmann equation in the semiclassical theory
and is responsible for regulating the divergent DC conductivity in the clean limit,
is incorporated into the response theory by regulating the divergent DC conductivity on the same footing.
Specifically, in the linear order of the electric field, we show that both formulations
can give the linear Drude current, intrinsic anomalous Hall current, and the displacement current,
where the last one arising from the time variation of the first-order \textit{positional shift} \cite{GaoY2014}
can only be induced by an AC electric field \cite{XiangDHE}.
Furthermore, in the second order of the electric field,
we show that the familiar second-order current responses,
including the nonlinear Drude current and the nonlinear Hall current,
which can be driven by Berry curvature dipole or quantum metric dipole,
can be derived by both formulations.
Importantly, we show that the intrinsic nonlinear longitudinal current,
as predicted by the response theory or in a similar manner,
can also be reproduced by the semiclassical theory,
particularly by considering the field-induced group velocity arising from the
second-order wavepacket energy \cite{XiaoC2ndenergy, XiaoCnormal}.
We wish to remark that the intrinsic longitudinal current due to the group velocity
in the semiclassical theory has been suppressed by
replacing $f_n(\epsilon_n)$ with $f_n(\bar{\epsilon}_n)$ \cite{GaoY2014},
where $\epsilon_n$ includes the second-order wavepacket energy correction.
Finally, with both formulations, we derive the extrinsic and intrinsic nonlinear displacement currents
unique to the AC electric field, which are overlooked previously and
arise from the time variation of the first-order and the second-order \textit{positional shift}s \cite{GaoY2014, xiang},
respectively.
Interestingly, we find that the extrinsic nonlinear displacement current features the same
quantum geometric origin with the intrinsic nonlinear Hall and longitudinal currents
but scales with a dimensionless constant $\omega \tau$, 
where $\omega$ stands for the driving frequency.
As a result, both the intrinsic nonlinear Hall and longitudinal currents
can be largely enhanced by the extrinsic nonlinear displacement current,
particularly under an AC electric field with $\omega \tau \gg 1$.
In addition, we note that the intrinsic nonlinear displacement conductivity 
features the $\mathcal{T}$-even ($\mathcal{T}$, time reversal) property
so that this nonlinear current can coexist with the $\mathcal{T}$-even
Berry curvature dipole driven nonlinear Hall effect under an AC electric field.

On top of the equivalence between both formulations,
(i) we suggest that the energy used in the equilibrium Fermi distribution
in the semiclassical theory is the unperturbed band energy
since both formulations are assumed to give the same intrinsic responses; 
(ii) we unify the intrinsic second-order nonlinear longitudinal current responses
calculated with different approaches;
(iii) we argue that the scheme of introducing relaxation in the response theory
by modifying the Liouville equation for the density matrix needs to be reconsidered
since Eq.(\ref{res1}) can give a linear extrinsic Hall current in $\mathcal{T}$-invariant systems,
as forbidden by the Onsager reciprocal relation \cite{Nagaosa2010, XiaoCnc, Onsager}.

The remainder of this paper is organized as follows.
In section \ref{ACsemitheory},
we develop the second-order AC semiclassical theory with the wavepacket method.
In section \ref{currentdensity},
we derive the current density up to the second order of the electric field with the extended
AC semiclassical theory and the response theory, respectively.
In section \ref{equivalence}, we explicitly show the
equivalence of the current density derived by the AC semiclassical theory and the response theory
up to the second order of the electric field, for both the zero-frequency and finite-frequency scenarios.
Based on this equivalence, in section \ref{consequence}
we clarify the inconsistent overlap between both formulations
discussed above and address the challenges confronted in each theory.
Furthermore, in section \ref{ACcorrection},
we discuss the nonlinear displacement currents unique to the AC transport,
which are derived by both formulations and
contain both intrinsic and extrinsic contributions.
In section \ref{summary}, we summarize the main results
and briefly outline the futural direction.
Finally, an Appendix \ref{appendix} is given.

\bigskip
\section{AC semiclassical theory}
\label{ACsemitheory}

In this section, we extend the AC semiclassical theory under a time-dependent and uniform electric field
to the second order of the driving field.
In detail, we construct the wavepacket accurate up to the second order of the AC electric field
and then evaluate the position center for the wavepacket,
where the key expressions for the \textit{positional shift}
up to the second order of the AC electric field are obtained.
In addition, we derive the equations of motions
for the wavepacket centers (including momentum and position)
find that the electric field can also induce a correction
to the mixed Berry curvature that is solely decided by the time variation of the second-order \textit{positional shift}.
Finally, we estimate the wavepacket energy accurate up to the second order of the AC electric field
and we find that this energy correction can also be expressed by the first-order \textit{positional shift},
like the extended DC semiclassical theory.

\bigskip
\subsection{The wavepacket accurate up to the second order}

Following the spirit of the semiclassical wavepacket theory \cite{Xiao2010, GaoY2014},
we focus on the $n$th band to construct the wavepacket as follows ($e=\hbar=1$):
\begin{align}
|W\rangle= \int_{\vect{p}} e^{i\vect{p}\cdot\vect{r}}
\left( C_{n} (\vect{p}) |u_{n}\rangle + \sum_{l \neq n} \bar{C}_{l} (\vect{p})
|u_{l} \rangle \right),
\label{wavepacket}
\end{align}
where $\int_{\vect{p}} \equiv \int d\vect{p}/(2\pi)^d$ with $d$ the system dimension,
$|u_n\rangle \equiv |u_n(\vect{p})\rangle$ is periodic part of the Bloch state, $C_n(\vect{p})$
is the zero-order amplitude
and $\bar{C}_{l}(\vect{p})$ stands for the correction
(indicated with a bar above $C_l$) induced by the external electric field
(Throughout this work, we will only consider the electric field).
Here we assume that \cite{Xiao2010}
\begin{align}
|C_n(\vect{p})|^2=\delta(\vect{p}-\vect{p}_c)
\label{normalize}
\end{align}
with $\vect{p}_c$ the momentum center
to normalize the wavepacket up to the first order \cite{Xiao2010}. Note that
Eq.(\ref{normalize}) is useful in evaluating various quantities related to the wavepacket.

As the superposition of Bloch states,
the wavepacket obeys the time-dependent Schr\"{o}dinger equation
\begin{align}
i\partial_t | W \rangle = H|W\rangle ,
\label{SE}
\end{align}
where $\partial_t=\partial/\partial t$, $H=H_0+H_1$ with $H_0$ the periodic crystal Hamiltonian,
which gives $H_0|u_n\rangle=\epsilon_n|u_n\rangle$,
and $H_1=\vect{E}(t)\cdot\vect{r}$ the perturbed Hamiltonian due to the applied AC electric field $\vect{E}(t)$.
Particularly, $\vect{E}(t)=\sum_{\omega_1}\vect{E} e^{-i(\omega_1+i\xi)t}$ with $\omega_1=\pm \omega$,
which is adiabatically turned on due to $\xi \rightarrow 0^{+}$.
By multiplying $\langle u_m(\vect{p})|e^{-i\vect{p}\cdot\vect{r}}$ on both sides of Eq.(\ref{SE}), we find
\begin{widetext}
\begin{align}
&\int_{\vect{p}'} e^{i(\vect{p}'-\vect{p})\cdot\vect{r}}
\left(
i\dot{C}_n(\vect{p}')\langle u_m|u_n' \rangle
+
\sum_{l \neq n}
i \dot{\bar{C}}_l(\vect{p}') \langle u_m|u_l' \rangle
\right)
=
\int_{\vect{p}'}
e^{i(\vect{p}'-\vect{p})\cdot\vect{r}}
\left(
C_n(\vect{p}')\langle u_m|H_0|u_n' \rangle
+
\sum_{l \neq n}
\bar{C}_l(\vect{p}') \langle u_m|H_0 |u_l'\rangle
\right)
\nonumber \\
& \qquad \qquad \qquad \qquad \qquad \qquad \qquad \qquad
+
\int_{\vect{p}'}
e^{i(\vect{p}'-\vect{p})\cdot\vect{r}}
\left(
C_n(\vect{p}')\langle u_m|H_1|u_n' \rangle
+
\sum_{l \neq n}
\bar{C}_l(\vect{p}')
\langle u_m|H_1|u_l' \rangle
\right),
\label{SE1}
\end{align}
\end{widetext}
where $\dot{C}_n=\partial_t C_n$ and $|u_n'\rangle=|u_n(\vect{p}')\rangle$.
At this stage, using the orthogonal relation between Bloch states
\begin{align}
\langle u_m|
e^{i(\vect{p}'-\vect{p})\cdot\vect{r}}
|
u_n'\rangle
=
\delta(\vect{p}'-\vect{p})\delta_{mn}
\end{align}
and the Bloch representation of position operator $\vect{r}$ \cite{Sipe1995, Sipe2000}
\begin{align}
\langle u_m|
e^{i(\vect{p}'-\vect{p})\cdot\vect{r}}
\vect{r}
|u_n'\rangle
=
\left[ \delta_{mn} i\partial_{\vect{p}} + \vect{\mathcal{A}}_{mn} \right] \delta(\vect{p}-\vect{p}'),
\end{align}
where $\partial_{\vect{p}} \equiv \partial/\partial \vect{p}$ and
$\vect{\mathcal{A}}_{nm} \equiv \langle u_n | i\partial_{\vect{p}} |u_m\rangle$,
when $m=n$ Eq.(\ref{SE1}) becomes \cite{footnote0}
\begin{align}
i \dot{C}_n  = \epsilon_n C_n
+
\vect{E}\cdot \left( i \partial_{\vect{p}} + \vect{\mathcal{A}}_n \right) C_n
+
\sum_{l} \bar{C}_l \vect{E} \cdot \vect{r}_{nl},
\label{Cn}
\end{align}
where $\vect{\mathcal{A}}_n \equiv \vect{\mathcal{A}}_{nn}$
is the intraband Berry connection and
$\vect{r}_{nl} \equiv \vect{\mathcal{A}}_{nl}$ for $n \neq l$
the interband Berry connection.
However, when $m \neq n$ Eq.(\ref{SE1}) reduces into
\begin{align}
i \dot{\bar{C}}_m = \epsilon_m \bar{C}_m
+
\vect{E} \cdot \vect{r}_{mn} C_n
&+
\vect{E} \cdot \left( i \partial_{\vect{p}} + \vect{\mathcal{A}}_m\right) \bar{C}_m
\nonumber \\
&+
\sum_{l} \bar{C}_l \vect{E} \cdot \vect{r}_{ml},
\label{barCm}
\end{align}
Eq.(\ref{barCm}), together with Eq.(\ref{Cn}), fully determines the unknown coefficient $\bar{C}_l$
in the wavepacket Eq.(\ref{wavepacket}).
Particularly, let $\bar{C}_m=M_{mn}C_n$, we find that
Eq.(\ref{barCm}) becomes
\begin{align}
&
i \dot{M}_{mn} C_n
+
M_{mn} i\dot{C}_n
=
\epsilon_m M_{mn} C_n
+
\vect{E} \cdot \vect{r}_{mn} C_n
\nonumber \\
&+
\vect{E} \cdot \left( i \partial_{\vect{p}} + \vect{\mathcal{A}}_m \right) M_{mn}C_n
+
\sum_{l}\vect{E} \cdot \vect{r}_{ml} M_{ln}C_n.
\end{align}
Then by subtituting Eq.(\ref{Cn}) into this equation and eliminating $C_n$, we finally obtain
\begin{align}
\dot{M}_{mn} + i \epsilon_{mn} M_{mn}
&=
-i \vect{E} \cdot \vect{r}_{mn}
+
\vect{E} \cdot \vect{\mathcal{D}}^{\vect{p}}_{mn} M_{mn}
\nonumber \\
&-
i
\sum_{l} \vect{E} \cdot (\vect{r}_{ml}-M_{mn}\vect{r}_{nl}) M_{ln},
\label{Mmn}
\end{align}
where
$\vect{\mathcal{D}}^{\vect{p}}_{mn} \equiv \partial_{\vect{p}}-i \left(\vect{\mathcal{A}}_m-\vect{\mathcal{A}}_n \right)$
is the $U(1)$ gauge-invariant derivative and $\epsilon_{mn}=\epsilon_{m} - \epsilon_n$ the energy difference.
Eq.(\ref{Mmn}) can be solved iteratively by writing $M_{mn}=\sum_i M_{mn}^{(i)}$ with $i \geq 1$
and $M_{mn}^{(i)} \propto E^{(i)}_\alpha$. For $i=1$, we find
\begin{align}
\dot{M}_{mn}^{(1)} +i \epsilon_{mn} M_{mn}^{(1)} = -i E_\alpha^{\omega_1} r^\alpha_{mn},
\label{Mmn1}
\end{align}
where $E_\alpha^{\omega_1}=E_\alpha e^{-i(\omega_1+i\xi)t}$ with $\omega_1=\pm \omega$
(Throughout this work, the summation over frequency is suppressed unless otherwise stated)
and the Einstein summation convention for the repeated Greek alphabets has been assumed,
and thereby
\begin{align}
M_{mn}^{(1)} &=
\dfrac{r_{mn}^\alpha E_\alpha^{\omega_1}}{\omega_1-\epsilon_{mn}+i\xi}.
\label{Mmn1}
\end{align}
Similarly, for $i=2$, we find:
\begin{align}
\dot{M}_{mn}^{(2)} &+ i \epsilon_{mn} M_{mn}^{(2)}
\nonumber \\
&=
E_\beta \mathcal{D}^\beta_{mn} M_{mn}^{(1)}
-
i\sum_{l} E_\beta^{\omega_2} r^\beta_{ml} M_{ln}^{(1)},
\label{Mmn2}
\end{align}
where $\mathcal{D}^\beta_{mn}=\partial_{\beta}-i(\mathcal{A}^\beta_{m}-\mathcal{A}_n^\beta)$
with $\partial_\beta=\partial/\partial p_\beta$. By solving Eq.(\ref{Mmn2}) we obtain
\begin{align}
M_{mn}^{(2)}
&=
i\mathcal{D}^\beta_{mn}
\left(
\dfrac{r^\alpha_{mn}}{\omega_1-\epsilon_{mn}+i\xi}
\right)
\dfrac{E_\alpha^{\omega_1} E_\beta^{\omega_2}}{\omega_{\Sigma}-\epsilon_{mn}+2i\xi}
\nonumber \\
&+
\sum_l
\dfrac{r^\beta_{ml}r^\alpha_{ln} E_\alpha^{\omega_1} E_\beta^{\omega_2}}
{(\omega_1-\epsilon_{ln}+i\xi)(\omega_{\Sigma}-\epsilon_{mn}+2i\xi)},
\label{Mmn2sol}
\end{align}
where $\omega_{\Sigma}=\omega_1+\omega_2$.

With $M_{mn}^{(1)}$ and $M_{mn}^{(2)}$,
the wavepacket accurate up to the second order of the electric field is given by
\begin{align}
|W\rangle
&=
\int_{\vect{p}} e^{i\vect{p}\cdot\vect{r}}
C_{n} (\vect{p})
\left[
|u_{n}\rangle
+
\sum_{m \neq n} \left( M_{mn}^{(1)} +  M_{mn}^{(2)} \right) |u_{m} \rangle
\right]
\nonumber \\
&\equiv
|W_0\rangle+|W_1\rangle+|W_2\rangle,
\label{wavepacket1}
\end{align}
where $\bar{C}_l=\sum_i C^{(i)}_m$ with $C_m^{(i)}=M_{mn}^{(i)}C_n$
has been used and $|W_i\rangle \propto E_\alpha^{(i)}$.

Note that the wavepacket will be used to evaluate the Lagrangian so we need to normalize it up to the second order
of the electric field,
which in fact has been carefully considered in the previous nonlinear semiclassical studies \cite{XiaoCnormal, GaoYPRB2019}.
Particularly, we have:
\begin{align}
|\bar{W}\rangle
&=
\dfrac{|W\rangle}{\sqrt{\langle W|W\rangle}}
=
\dfrac{|W\rangle}{\sqrt{1+\langle W_1|W_1\rangle+\mathcal{O} (E_\alpha^3)}}
\nonumber \\
&=
\left[ 1 + \delta + \mathcal{O} (E_\alpha^3) \right] |W\rangle,
\end{align}
where we have performed a Taylor expansion and obtained \cite{GaoYPRB2019}
\begin{align}
\delta
&\equiv
-\dfrac{1}{2}
\langle W_1|W_1\rangle
=
-\dfrac{1}{2}
\sum_m
M_{mn}^{(1)} (\vect{p}_c) M_{mn}^{(1)*} (\vect{p}_c),
\label{delta}
\end{align}
by using Eq.(\ref{normalize}) and Eq.(\ref{wavepacket1}).
Finally, the normalized AC wavepacket accurate up to the second order of the electric field
can be expressed as
\begin{align}
|\bar{W}\rangle
=
(1+\delta)|W_0\rangle+|W_1\rangle+|W_2\rangle,
\label{wavepacket2}
\end{align}
which constitutes the starting point to formulate the AC semiclassical theory
particularly accurate up to the second order of the electric field.

\subsection{The position center for the wavepacket}
With the constructed wavepacket $|\bar{W}\rangle$, the position center $\vect{r}_c$ defined by
$\vect{r}_c \equiv \langle \bar{W}|\vect{r}|\bar{W}\rangle$ is found to be
(see Appendix \ref{pcenter})
\begin{align}
\vect{r}_c
=
\partial_{\vect{p}_c}\gamma  + \vect{\mathcal{A}}_n  + \vect{a}_n^{(1)} + \vect{a}_n^{(2)} ,
\label{rcdef}
\end{align}
where $\gamma=-\text{Arg}(C_n)$ and we have suppressed the dependence on $\vect{p}_c$ for all quantities for brevity.
In addition, here $\vect{a}_n^{(i)}$ is the $i$th \textit{positional shift} given by
\begin{align}
a_n^{(1;\lambda)}
&=
2\text{Re}
\sum_{m}
\dfrac{r^\lambda_{nm}r^\alpha_{mn}}{\omega_1-\epsilon_{mn}+i\xi} E_\alpha^{\omega_1},
\label{an1}
\end{align}
and
\begin{align}
a_n^{(2;\lambda)}
&=
\text{Re}
\sum_{ml}
\dfrac{2r^\lambda_{nm}r^\beta_{ml}r^\alpha_{ln}}{(\omega_1-\epsilon_{ln}+i\xi)(\omega_{\Sigma}-\epsilon_{mn}+2i\xi)}
E_\alpha^{\omega_1} E_\beta^{\omega_2}
\nonumber \\
&+
\sum_{ml}
\dfrac{-r^\beta_{nl} r^\lambda_{lm} r^\alpha_{mn}}{(\omega_2-\epsilon_{nl}+i\xi)(\omega_1-\epsilon_{mn}+i\xi)}
E_\alpha^{\omega_1} E_\beta^{\omega_2}
\nonumber \\
&+
\text{Re}
\sum_m
i\mathcal{D}^\beta_{mn}
\left(
\dfrac{r^\alpha_{mn}}{\omega_1-\epsilon_{mn}+i\xi}
\right)
\dfrac{2r^\lambda_{nm}E_\alpha^{\omega_1} E_\beta^{\omega_2}}{\omega_{\Sigma}-\epsilon_{mn}+2i\xi}
\nonumber \\
&+
\text{Re}
\sum_{m}
i\mathcal{D}^\lambda_{nm}
\left(
\dfrac{r^\alpha_{nm}}{\omega_1-\epsilon_{nm}+i\xi}
\right)
\dfrac{r^\beta_{mn}E_\alpha^{\omega_1} E_\beta^{\omega_2}}{\omega_2-\epsilon_{mn}+i\xi},
\label{an2}
\end{align}
where again the Einstein summation convention for the repeated Greek alphabets is assumed
and the summation over frequency has been suppressed.
Note that under $U(1)$ gauge transformation $|u_n\rangle \rightarrow e^{i\phi_n}|u_n\rangle$,
$\mathcal{A}_n^\alpha \rightarrow \mathcal{A}_n^\alpha - \partial_\alpha \phi_n$,
$r^\alpha_{nm} \rightarrow e^{i(\phi_m-\phi_n)}r^\alpha_{nm}$,
it is easy to check that both $a_n^{(1;\lambda)}$ and $a_n^{(2;\lambda)}$
are gauge invariant, as they must be.

\subsection{The equations of motion}
With the normalized AC wavepacket $|\bar{W}\rangle$, the equations of motion for the wavepacket
centers $\vect{p}_c$ and $\vect{r}_c$ can also be derived. To that purpose,
we first evaluate the Lagrangian for the wavepacket, which is defined as:
\begin{align}
\mathcal{L}
\equiv
\langle \bar{W} | (i\partial_t - H) |\bar{W}\rangle
=
\langle W_0|i\partial_t|W_0\rangle - \bar{\epsilon}_n,
\end{align}
where $\bar{\epsilon}_n \equiv \langle W_0|i\partial_t|W_0\rangle - \langle \bar{W} | (i\partial_t - H) |\bar{W}\rangle$
is the wavepacket energy (accurate up to the second order of the electric field for our purpose)
that will be calculated in the subsequent subsection. Note that
\begin{align}
\langle W_0|i\partial_t|W_0\rangle =\partial_t \gamma(\vect{p}_c),
\label{W0tW0}
\end{align}
and therefore
\begin{align}
\mathcal{L} &= \partial_t \gamma - \bar{\epsilon}_n = \dfrac{d}{dt}\gamma - \partial_{\vect{p}_c} \gamma \cdot \dot{\vect{p}}_c
-
\bar{\epsilon}_n
\nonumber \\
&=
-\dot{\vect{k}}_c \cdot (\vect{r}_c - \vect{\mathcal{\bar{A}}}_n)
-
\bar{\epsilon}_n,
\end{align}
where we have used Eq.(\ref{rcdef}) and
defined $\vect{\mathcal{\bar{A}}}_n \equiv \vect{\mathcal{A}}_n+\vect{a}_n^{(1)}+\vect{a}_n^{(2)}$.
Here we have dropped the unimportant total time derivative $d\gamma/dt$ in the Lagrangian
and replaced $\vect{p}_c$ with $\vect{k}_c$ in the final result.
In addition, we have assumed that the eigenstates of Hamiltonian do not explicitly depend on time 
so that $\vect{\mathcal{A}}^t_n \equiv \langle u_n |\partial_t|u_n \rangle=0$ \cite{Xiao2010}.
Next, using the Euler-Lagrange equations,
\begin{align}
\dfrac{d}{dt} \dfrac{\partial \mathcal{L}}{\partial \dot{\vect{k}}_c}
=
\dfrac{\partial \mathcal{L}}{\partial \vect{k}_c},
\quad
\dfrac{d}{dt} \dfrac{\partial \mathcal{L}}{\partial \dot{\vect{r}}_c}
=
\dfrac{\partial \mathcal{L}}{\partial \vect{r}_c},
\end{align}
we find that
\begin{align}
\dot{\vect{r}} &= -\dot{\vect{k}} \times \vect{\bar{\Omega}}_n+\bar{\vect{v}}_n+\partial_t \vect{\mathcal{\bar{A}}}_n,
\label{rEOM}
\\
\dot{\vect{k}} &=-\vect{E}(t),
\label{kEOM}
\end{align}
where we have further suppressed the subscript $c$ on $\vect{r}_c$ and $\vect{k}_c$.
In addition, we have used $\partial \bar{\epsilon}_n/\partial \vect{r}_c=\vect{E}(t)$
from Eq. (\ref{energy})
and defined
$\partial \bar{\epsilon}_n/\partial \vect{k}_c \equiv \bar{\vect{v}}_n$
and
$\vect{\bar{\Omega}}_n \equiv \partial/\partial\vect{k}_c \times \vect{\mathcal{\bar{A}}}_n$.
Note that Eqs. (\ref{rEOM}-\ref{kEOM})
are formally the same as the first-order or linear AC semiclassical theory developed by
G. Sundaram and Q. Niu in Ref. [\onlinecite{Niu1999}].

We wish to remark that the equations of motion for the AC semiclassical theory is the same
as the DC semiclassical theory except for the last term in Eq.(\ref{rEOM}).
In fact, this term is only contributed by the time variation of \textit{positional shift}
since $\vect{\mathcal{A}}_n$ is time-independent, especially,
$\partial_t \vect{\mathcal{\bar{A}}}_n = \partial_t (\vect{a}_n^{(1)}+\vect{a}_n^{(2)})$.
Remarkably, this additional term will give the displacement current under the
time-dependent external electric field \cite{XiangDHE} and is usually overlooked even in the AC transport.
Interestingly, we note that the \textit{positional shift} also enters
the first term of Eq.(\ref{rEOM}), which gives the
anomalous velocity $\vect{E} \times \vect{\Omega}^{(i)}_n$
at the $(i+1)$th order of the electric field
since $\vect{\Omega}^{(i)}_n = \partial_{\vect{k}} \times \vect{a}_n^{(i)}$ with $i\geq 1$.
As a result, the first-order \textit{positional shift}
shows up only in the second-order DC semiclassical theory,
whereas the same \textit{positional shift} must be calculated in the first-order AC semiclassical theory.
In addition, the energy correction for the wavepacket particularly
at second order of the electric field in fact is determined by
the first-order \textit{positional shift}, as will be shown below.

\subsection{The wavepacket energy}

To complete the AC semiclassical theory particularly accurate up to the second order of the electric field,
we now derive the wavepacket energy $\bar{\epsilon}_n$ accurate up to the same order,
which determines the group velocity appeared in Eq.(\ref{rEOM}). In particular, we find
(see Appendix \ref{wpenergy}),
\begin{align}
\bar{\epsilon}_n
&\equiv
\langle W_0|i\partial_t|W_0\rangle - \langle \bar{W} | (i\partial_t - H) |\bar{W}\rangle
\nonumber \\
&=\epsilon_n+\vect{E}\cdot\vect{r}_c+\epsilon^{(2)}_n+\mathcal{O}(E_\alpha^3)
\label{energy}
\end{align}
where
\begin{align}
\epsilon_n^{(2)}
&\equiv
\text{Re}
\sum_{m}
\left[
\left(i\partial_t M_{mn}^{(1)}\right) M_{mn}^{(1)*}
+
\epsilon_{mn} M_{mn}^{(1)} M_{mn}^{(1)*}
\right]
\nonumber \\
&=
-\text{Re}\sum_{m}
\dfrac{(\omega_1+i\xi+\epsilon_{mn})r_{mn}^\alpha r^\beta_{nm} E_\alpha^{\omega_1}E_\beta^{\omega_2}}
{(\omega_1-\epsilon_{mn}+i\xi)(\omega_2-\epsilon_{nm}+i\xi)}
\nonumber \\
&=
-\dfrac{1}{2}
a_n^{(1;\alpha)} E_\alpha^{\omega_1},
\label{secondenergy}
\end{align}
where we have used Eq.(\ref{Mmn1}), Eq.(\ref{an1}), and $M_{mn}^{(1)*}=-M_{nm}^{(1)}$.
We stress that $\vect{E}\cdot\vect{r}_c$ does not enter the semiclassical equations of motion
since $\vect{r}_c$ and $\vect{k}_c$ are independent variables in the Lagrangian \cite{Xiao2010}.
Note that Eq.(\ref{secondenergy}) in the DC limit
recovers the second-order energy correction
obtained with the extended DC semiclassical theory \cite{XiaoC2ndenergy, XiaoCnormal}, as expected.
Eqs.(\ref{rEOM}-\ref{kEOM}), combining with the Boltzmann equation,
can be used to calculate the AC current density,
as will be discussed in the following section.

\section{Current density}
\label{currentdensity}

To clearly show the equivalence between the AC semiclassical theory
and the response theory, in this section we first derive the current density
(or the current density conductivity tensor) calculated
from both the AC semiclassical theory formulated above and the response theory
at the length gauge \cite{Sipe1995, Sipe2000}, respectively.

\subsection{The current density from the AC semiclassical theory}
\label{semicurrentdensity}

Armed with the AC semiclassical theory accurate up to the second order of the electric field,
the current density defined by \cite{Xiao2010}
\begin{align}
J_\lambda = \sum_n \int_k \bar{f}_n \dot{r}_n^\lambda,
\label{current}
\end{align}
can be easily calculated by solving Eqs.(\ref{rEOM}-\ref{kEOM}) for the semiclassical velocity $\dot{r}_n^\lambda$
as well as Eq.(\ref{boltz0}) for the nonequilibrium distribution function $\bar{f}_n$.
Particularly, up to the second order of the electric field, we find
\begin{align}
\dot{r}_n^\lambda
=
\left( v_n^\lambda+v^{(2;\lambda)}_n \right)
&+
\left(
\Omega^{\lambda\alpha}_n
+
\Omega^{(1;\lambda\alpha)}_n
\right)
E_\alpha^{\omega_1}
\nonumber \\
&+
\partial_t
\left(
a_n^{(1;\lambda)}
+
a_n^{(2;\lambda)}
\right),
\label{dotr}
\end{align}
where $v_n^{(2;\lambda)}=\partial_\lambda \epsilon_n^{(2)}$ with $\epsilon_n^{(2)}$ calculated in Eq.(\ref{secondenergy}),
$\Omega^{\lambda\alpha}_n=i\sum_{m}(r^\lambda_{nm}r^\alpha_{mn}-r^\alpha_{nm}r^\lambda_{mn})$
is the conventional Berry curvature, and
$\Omega_n^{(1;\lambda\alpha)}=\partial_\lambda a_n^{(1;\alpha)}-\partial_\alpha a_n^{(1;\lambda)}$
is the field-induced Berry curvature due to the first-order \textit{positional shift} calculated in Eq.(\ref{an1}).
In addition, by solving the Boltzmann equation we find (see Appendix \ref{neqsol})
\begin{align}
f^{(1)}_n &=
\dfrac{i \partial_\alpha f_n E_\alpha^{\omega_1}}{\omega_1 + i (\eta+\xi)},
\label{neq1}
\\
f^{(2)}_n &=
\dfrac{- \partial_{\alpha\beta}^2 f_n E_\alpha^{\omega_1} E_\beta^{\omega_2}}
{\left[\omega_1+i(\eta+\xi)\right[\left[\omega_\Sigma+i(\eta+2\xi)\right]},
\label{neq2}
\end{align}
where $\eta \equiv 1/\tau$ is the scattering rate.
Then by inserting Eqs.(\ref{dotr}-\ref{neq2}) into Eq.(\ref{current}),
we immediately obtain the first-order current density
\begin{align}
J_\lambda^{(1)} & \equiv J_\lambda^{(1;1)} + J_\lambda^{(1;0)} +  J_\lambda^{(1;\bar{0})}
\end{align}
with
\begin{align}
J_\lambda^{(1;1)}
&=
\sum_n \int_k
f_n^{(1)} v_n^\lambda
=
\sum_n \int_k
\dfrac{i \partial_\alpha f_n v_n^\lambda E_\alpha^{\omega_1}}{\omega_1 + i (\eta+\xi)},
\label{J11}
\\
J_\lambda^{(1;0)}
&=
\sum_n \int_k
f_n
\Omega_n^{\lambda\alpha} E_\alpha^{\omega_1},
\label{J10}
\\
J_\lambda^{(1;\bar{0})}
&=
\sum_n \int_k
f_n
\partial_t a_n^{(1;\lambda)},
\label{J1bar0}
\end{align}
where the bar over $n$ indicates that the current density
vanishes in the DC limit,
and the second-order current density
\begin{align}
J_\lambda^{(2)} \equiv J_\lambda^{(2;0)}+J_\lambda^{(2;1)}+J_\lambda^{(2;2)}
\end{align}
with
\begin{align}
J_\lambda^{(2;0)} &=
\sum_n \int_k f_n
\left(
v_n^{(2;\lambda)} +
\Omega^{(1;\lambda\alpha)}_nE_\alpha^{\omega_1}
+
\partial_t a_n^{(2;\lambda)}
\right),
\label{J20}
\\
J_\lambda^{(2;1)} &=
\sum_n \int_k f_n^{(1)}
\left(
\Omega^{\lambda\alpha}_n E_\alpha^{\omega_1}
+
\partial_t a_n^{(1;\lambda)}
\right),
\label{J21}
\\
J_\lambda^{(2;2)} &= \sum_n \int_k f_n^{(2)}v_n^\lambda
\nonumber \\
&=
\sum_n \int_k
\dfrac{- \partial_{\alpha\beta}^2 f_n v_n^\lambda E_\alpha^{\omega_1} E_\beta^{\omega_2}}
{\left[\omega_1+i(\eta+\xi)\right[\left[\omega_\Sigma+i(\eta+2\xi)\right]}.
\label{J22}
\end{align}
Note that both $J_\lambda^{(2;0)}$ and $J_\lambda^{(2;1)}$ contains
contributions that disappear in the DC limit, which are defined as
\begin{align}
J_\lambda^{(2;\bar{0})} &\equiv \sum_n \int_k f_n \partial_t a_n^{(2;\lambda)},
\label{nonlineardispint}
\\
J_\lambda^{(2;\bar{1})} &\equiv \sum_n \int_k f_n^{(1)} \partial_t a_n^{(1;\lambda)}.
\end{align}

In addition to the AC semiclassical theory, the current density
can also be calculated by the response theory through solving the quantum Liouville
equation for the nonequilibrium density matrix, which is summarized below.

\subsection{The current density from the response theory}

In subsection \ref{semicurrentdensity}, we have expressed the results of AC semiclassical theory
in terms of the current density; to make a clear notation distinction,
below we will use the conductivity tensor of the current density
to express the results given by the response theory.
In particular, under the framework of the length-gauge response theory,
we find that the linear conductivity can be decomposed as
(see Appendix \ref{responsesol})
\begin{align}
\sigma^{(1)}_{\lambda\alpha}
&=
\sigma^{(1;1)}_{\lambda\alpha}
+
\sigma^{(1;0)}_{\lambda\alpha}
+
\sigma^{(1;\bar{0})}_{\lambda\alpha},
\label{sigma1}
\end{align}
where
\begin{align}
\sigma^{(1;1)}_{\lambda\alpha}
&=
\sum_{n} \int_k
\dfrac{i\partial_{\alpha}f_n}{\omega_1+i\xi}v_n^\lambda ,
\label{sigmaDrude} \\
\sigma^{(1;0)}_{\lambda\alpha}
&=
\sum_{n} \int_k f_n \Omega_{n}^{\lambda\alpha},
\label{sigmaIAHE} \\
\sigma^{(1;\bar{0})}_{\lambda\alpha}
&=
\sum_{mn} \int_k
\dfrac{-i(\omega_1+i\xi)r^\lambda_{nm}r^\alpha_{mn}f_{nm}}{\omega_1-\epsilon_{mn}+i\xi},
\label{sigmaDisp}
\end{align}
which contains three contributions same as the AC semiclassical theory.

Furthermore, in the second order of the electric field,
the conductivity tensor evaluated with the response theory
can be decomposed as \cite{Hipolito} (see Appendix \ref{responsesol}):
\begin{align}
\sigma_{\lambda\alpha\beta}^{(2)}
=
\sigma_{\lambda\alpha\beta}^{(2,ee)}
+
\sigma_{\lambda\alpha\beta}^{(2,ei)}
+
\sigma_{\lambda\alpha\beta}^{(2,ie)}
+
\sigma_{\lambda\alpha\beta}^{(2,ii)},
\label{sigma2}
\end{align}
where
\begin{align}
\sigma_{\lambda\alpha\beta}^{(2,ee)}
&=
\sum_{mnl} \int_k
\mathcal{V}^{\lambda}_{nm}
\left(
\dfrac{r^\beta_{ml} r^\alpha_{ln}f_{nl}}{\omega_1-\epsilon_{ln}+i\xi}
-
\dfrac{f_{lm}r^\alpha_{ml}r^\beta_{ln}}{\omega_1 -\epsilon_{ml}+i\xi}
\right),
\label{sigmaee} \\
\sigma_{\lambda\alpha\beta}^{(2,ei)}
&=
\sum_{mn} \int_k
\mathcal{V}^{\lambda}_{nm}
\left(\dfrac{i r^\beta_{mn}\partial_{\alpha}f_{nm}}{\omega_1+i\xi}\right),
\label{sigmaei}
\\
\sigma_{\lambda\alpha\beta}^{(2,ie)}
&=
\sum_{mn} \int_k
\mathcal{V}^{\lambda}_{nm}
\mathcal{D}_{mn}^\beta
\left(\dfrac{i f_{nm}r^\alpha_{mn}}{\omega_1-\epsilon_{mn}+i\xi}
\right),
\label{sigmaie}
\\
\sigma_{\lambda\alpha\beta}^{(2,ii)}
&=
\sum_{n} \int_k
\dfrac{-v_n^\lambda \partial^2_{\alpha \beta} f_n}{(\omega_1+ i\xi)(\omega_\Sigma+2i\xi)},
\label{sigmaii}
\end{align}
where $f_{nl}=f_n-f_l$, $\omega_\Sigma=\omega_1+\omega_2$, and
\begin{align}
\mathcal{V}^{\lambda}_{nm}
=
\dfrac{v^\lambda_{nm} }{\omega_\Sigma - \epsilon_{mn} + 2i\xi}.
\end{align}
Similar to the second-order current density calculated using the AC semiclassical theory,
there exists many contributions at this order.

\begin{figure}[t!]
\centering
\includegraphics[width=0.85\columnwidth]{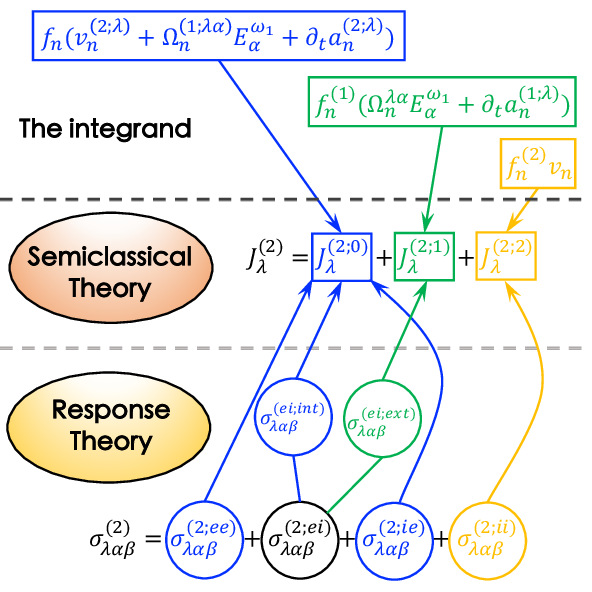}
\caption{
\label{FIG1}
The schematic illustration of the equivalence between the AC semiclassical theory
and the response theory. Here the middle panel displays the second-order current
density given by the AC semiclassical theory, where the integrand for each term
is placed into the top panel, and the bottom panel shows the second-order nonlinear conductivity
given by the length-gauge response theory. The arrows with different colors are used
to indicate the equivalence between both formulations.
We stress that $f_n\partial_t a_n^{2;\lambda}$ and $f_n^{(1)}\partial_t a_n^{1;\lambda}$
gives the intrinsic and extrinsic nonlinear displacement currents under an AC electric field,
respectively, as first proposed in this work.}
\end{figure}

\section{Equivalence between the AC semiclassical theory and the response theory}
\label{equivalence}

Once the current density is obtained by both the AC semiclassical theory and the response theory,
we are ready to show that the current density given by both formulations
are equivalent. To that purpose, we note that the relaxation (indicated by $\tau$ or $\eta=1/\tau$)
has been explicitly included into the semiclassical currents
(particularly in $J_\lambda^{(1;1)}$, $J_\lambda^{(2;1)}$ and $J_\lambda^{(2;2)}$)
by way of solving the Boltzman equation and thereby corresponds to the extrisnic contributions,
while the conductivities given by the response theory does not include any relaxation effects.
However, by adopting the clean limit ($\tau \rightarrow \infty$ or $\eta \rightarrow 0$), 
we show that all these extrinsic semiclassical results
can be exactly reproduced by the response theory, in which they show
the same $1/\xi$-type divergent behavior under the DC limit ($\omega \rightarrow 0$)
(Note that $\xi \rightarrow 0^+$). Recall that this divergence appeared in the semiclassical theory can be properly
regulated by taking a finite $\tau$ due to the presence of relaxation,
as a result, by replacing $\xi$ with $n/\tau$ \cite{Yanunify} in the
corresponding expressions given by the response theory,
we find that the equivalence between both formulations particularly in the clean limit
can be inherited in the presence of relaxation.
In addition to the extrinsic contributions, we further show that the intrinsic contributions
(including $J_\lambda^{(1;0)}$ and $J_\lambda^{(2;0)}$ which are independent of $\tau$)
given by the semiclassical theory can also be exactly reproduced by the response theory, 
where $\xi \rightarrow 0$ can no longer lead to a divergent conductivity in the DC limit
so that there is no need for regulating.

\subsection{The first order}
At the linear order, by comparing Eqs. (\ref{J11}-\ref{J10}) given by the AC semiclassical theory
with Eqs. (\ref{sigmaDrude}-\ref{sigmaIAHE}) given by the response theory, it is easy to find that
\begin{align}
\sigma_{\lambda\alpha}^{(1;1)}E_\alpha^{\omega_1} = \lim_{\eta \rightarrow 0} J_\lambda^{(1;1)},
\quad
\sigma_{\lambda\alpha}^{(1;0)}E_\alpha^{\omega_1} = J_\lambda^{(1;0)},
\label{equifirst1}
\end{align}
which gives the well-known linear Drude current and intrinsic anomalous Hall current, respectively.
Particularly, in the DC limit, we find
\begin{align}
\lim_{\omega_1 \rightarrow 0}
\lim_{\eta \rightarrow 0}
J_\lambda^{(1;1)}
&=
\dfrac{1}{\xi}
\sum_n \int_k \partial_\alpha f_n v_n^\lambda E_\alpha
\nonumber \\
&=
\lim_{\omega_1 \rightarrow 0}
\sigma^{(1;1)}_{\lambda\alpha}E_\alpha^{\omega_1}
\label{eq62}.
\end{align}
Note that this divergence can be properly regulated by taking a finite $\tau$ particularly in the semiclassical result,
as result, by replacing $\xi$ by $1/\tau$ in the expression given by the response theory, we have
\begin{align}
\lim_{\omega_1 \rightarrow 0}
\lim_{\xi \rightarrow 0}
J_\lambda^{(1;1)}
&=
\tau
\sum_n \int_k \partial_\alpha f_n v_n^\lambda E_\alpha
\nonumber \\
&=
\lim_{\omega_1 \rightarrow 0}
\sigma^{(1;1)}_{\lambda\alpha}(\xi \rightarrow 1/\tau)E_\alpha^{\omega_1}.
\label{equifirst11}
\end{align}
By doing this, we find that both formulations still give the same expression
in the presence of relaxation.
In addition, we have
\begin{align}
\lim_{\omega_1 \rightarrow 0} J_\lambda^{(1;0)}
=
\sum_n \int_k f_n \Omega_n^{\lambda\alpha}E_\alpha
=
\lim_{\omega_1 \rightarrow 0}
\sigma_{\lambda\alpha}^{(1;0)} E_\alpha^{\omega_1},
\label{equifirst2}
\end{align}
which stands for the intrinsic contribution
and therefore there is no need for regulating even in the DC limit.

Furthermore, by inserting Eq.(\ref{an1}) into Eq.(\ref{J1bar0}), we find
\begin{align}
J_\lambda^{(1;\bar{0})}
&=
\sum_{nm}
\int_k f_n
\left(
\dfrac{r^\lambda_{nm}r^\alpha_{mn}\partial_t E_\alpha^{\omega_1}}{\omega_1-\epsilon_{mn}+i\xi}
+
\dfrac{r^\lambda_{mn}r^\alpha_{nm} \partial_t  E_\alpha^{-\omega_1}}{\omega_1-\epsilon_{mn}-i\xi}
\right)
\nonumber \\
&=
\sum_{nm}
\int_k f_{nm}
\dfrac{-i(\omega_1+i\xi)r^\lambda_{nm}r^\alpha_{mn}}{\omega_1-\epsilon_{mn}+i\xi} E_\alpha^{\omega_1}
\nonumber \\
&= \sigma_{\lambda\alpha}^{(1;\bar{0})}E_\alpha^{\omega_1},
\label{equifirst3}
\end{align}
where we have used $\sum_{\omega_1=\pm \omega}h(\omega_1) = \sum_{\omega_1=\pm \omega} h(-\omega_1)$
for an arbitrary function $h(\omega_1)$ and interchanged $m$ and $n$ to obtain the second line.
We wish to mention that Eq.(\ref{J1bar0}) or the equivalent
$\sigma_{\lambda\alpha}^{(1;\bar{0})}E_\alpha^{\omega_1}$
gives the linear intrinsic displacement current under an AC electric field (vanishes in the DC limit),
which is intimately related to the quantum metric as
recently discussed in Ref. [\onlinecite{XiangDHE}].
With Eqs. (\ref{equifirst1}), (\ref{equifirst11}), (\ref{equifirst2}) and (\ref{equifirst3}),
we conclude that the AC semiclassical current at the linear order of the electric field
is equivalent to that given by the response theory.

\subsection{The second order: the DC case}

Since the AC semiclassical theory and the response theory give many terms
at the second order of the electric field, therefore,
before considering the general situation we first
show the equivalence between both formulations in the DC limit,
where the contributions due to the time variation of \textit{positional shift}
in Eqs. (\ref{J20}) and (\ref{J21}) given by the AC semiclassical theory do not appear.

In the DC limit, by investigating the zero-frequency divergence of both approaches,
we first note that Eq.(\ref{J22}) obtained with the AC semiclassical theory under the clean limit ($\eta \rightarrow 0$)
is equivalent to Eq.(\ref{sigmaii}) obtained with the response theory.
Explicitly, we find
\begin{align}
\lim_{\omega_i \rightarrow 0} \lim_{\eta \rightarrow 0}
J^{(2;2)}_\lambda
&=
\dfrac{1}{\xi}
\times
\dfrac{1}{2\xi}
\sum_n \int_k v_n^\lambda \partial^2_{\alpha\beta} f_n E_\alpha E_\beta
\nonumber \\
&=
\lim_{\omega_i \rightarrow 0}
\sigma^{(2;ii)}_{\lambda\alpha\beta}E_\alpha^{\omega_1} E_\beta^{\omega_2}.
\label{Drude2-0}
\end{align}
Similarly, this divergence can also be properly regulated by taking a finite $\tau$ in the semiclassical theory,
as result, by replacing $1/(2\xi^2)$ by $\tau^2$ \cite{Yanunify} in the expression given by the response theory,
we arrive at
\begin{align}
\lim_{\omega_i \rightarrow 0} \lim_{\xi \rightarrow 0}
J^{(2;2)}_\lambda
&=
\tau^2
\sum_n \int_k v_n^\lambda \partial^2_{\alpha\beta} f_n E_\alpha E_\beta
\nonumber \\
&=
\lim_{\omega_i \rightarrow 0}
\sigma^{(2;ii)}_{\lambda\alpha\beta}(2\xi^2 \rightarrow 1/\tau^2)E_\alpha^{\omega_1} E_\beta^{\omega_2}.
\label{Drude2-00}
\end{align}
By doing this, we find that both formulations still give the same result
in the presence of relaxation,
which is nothing but the familiar nonlinear Drude current
which quadratically depends on the relaxation time $\tau$.

Following the same strategy, we next show that the semiclassical current $J^{(2;1)}_\lambda$ given by Eq.(\ref{J21})
in the DC limit can be given by the nonlinear conductivity $\sigma^{(2,ei)}_{\lambda\alpha\beta}$
given by Eq.(\ref{sigmaei}) in the response theory.
Particularly, by inserting Eq.(\ref{neq1}) into Eq.(\ref{J21}),
we first find that $J^{(2;1)}_\lambda$ in the DC limit becomes
\begin{align}
\lim_{\omega_i \rightarrow 0}
\lim_{\eta \rightarrow 0}
J^{(2;1)}_{\lambda}
&=
\frac{1}{\xi}
\sum_{n} \int_k \partial_\beta f_n \Omega_{n}^{\lambda\alpha}E_\alpha E_\beta,
\end{align}
On the other hand,
using $v^\lambda_{nm}=i\epsilon_{nm}r^\lambda_{nm}$ ($m \neq n$) and
\begin{align}
&\dfrac{\epsilon_{mn}}{(\omega_\Sigma-\epsilon_{mn}+2i\xi)(\omega_1+i\xi)}
\nonumber \\
&=
\dfrac{\epsilon_{mn}}{\omega_2-\epsilon_{mn}+i\xi}
\left(
\dfrac{1}{\omega_1+i\xi}
-
\dfrac{1}{\omega_\Sigma-\epsilon_{mn}+2i\xi}
\right),
\label{factorization}
\end{align}
we find that Eq.(\ref{sigmaei}) in the DC limit can be decomposed as
$
\sigma^{(2;ei)}_{\lambda\alpha\beta}
=
\sigma^{(ei;ext)}_{\lambda\alpha\beta}+\sigma^{(ei;int)}_{\lambda\alpha\beta}
$,
where
\begin{align}
\lim_{\omega_i \rightarrow 0} \sigma^{(ei;ext)}_{\lambda\alpha\beta}
&=
\frac{1}{\xi}
\sum_{mn} \int_k i r^\lambda_{nm} r^\beta_{mn} \partial_\alpha f_{nm},
\label{sigmaeiext0}
\\
\lim_{\omega_i \rightarrow 0} \sigma^{(ei;int)}_{\lambda\alpha\beta}
&=
\sum_{mn} \int_k \dfrac{r^\lambda_{nm}r^\beta_{mn}\partial_\alpha f_{nm}}{-\epsilon_{mn}}.
\label{sigmaeiint0}
\end{align}
Here the superscript $ext$ ($int$) stands for the extrinsic (intrinsic) contribution.
By interchanging $m$ and $n$ of Eq.(\ref{sigmaeiext0}), it is easy to show that
\begin{align}
\lim_{\omega_i \rightarrow 0} \sigma^{(ei;ext)}_{\lambda\alpha\beta}E_\alpha^{\omega_1} E_\beta^{\omega_2}
&=
\frac{1}{\xi}
\sum_{n} \int_k \partial_\alpha f_n \Omega^{\lambda\beta}_n E_\alpha E_\beta
\nonumber \\
&=
\lim_{\omega_i \rightarrow 0} \lim_{\eta \rightarrow 0} J^{(2;1)}_{\lambda}.
\end{align}
Beyond the clean limit or when the relaxation is present,
this divergent conductivity can be naturally regulated by taking a finite $\tau$
in the semiclassical expression, as a result, also by replacing $1/\xi$ by $\tau$, we have
\begin{align}
\lim_{\omega_i \rightarrow 0}
\sigma^{(ei;ext)}_{\lambda\alpha\beta}(\xi \rightarrow 1/\tau)E_\alpha^{\omega_1} E_\beta^{\omega_2}
&=
\tau
\sum_{n} \int_k \partial_\alpha f_n \Omega^{\lambda\beta}_n E_\alpha E_\beta
\nonumber \\
&=
\lim_{\omega_i \rightarrow 0} \lim_{\xi \rightarrow 0} J^{(2;1)}_{\lambda},
\end{align}
which is nothing but the well-known extrinsic second-order nonlinear Hall current
driven by the Berry curvature dipole \cite{BCD, JEMoore2010}.

Finally, we show that the intrinsic conductivity Eq.(\ref{sigmaeiint0})
together with the remaining conductivities (which can only contribute to the intrinsic current, as will be clear below)
in the response theory, namely, Eq.(\ref{sigmaee}) and Eq.(\ref{sigmaie}),
is equivalent to the intrinsic current given by Eq.(\ref{J20}) in the AC semiclassical theory,
particularly in the DC limit.
Explicitly, we will show:
\begin{align}
\lim_{\omega_i \rightarrow 0} 
J^{(2;0)}_{\lambda}
&=
\lim_{\omega_i \rightarrow 0}
\left[
\sigma^{(2; ee)}_{\lambda\alpha\beta}
+
\sigma^{(2; ie)}_{\lambda\alpha\beta}
+
\sigma^{(ei;int)}_{\lambda\alpha\beta}
\right]
E_\alpha^{\omega_1} E_\beta^{\omega_2}.
\end{align}
To that purpose, by inserting Eq.(\ref{secondenergy}) and
$\Omega_n^{(1;\lambda\alpha)}=\partial_\lambda a_n^{(1;\alpha)}-\partial_\alpha a_n^{(1;\lambda)}$
into Eq.(\ref{J20}), we find that $J_\lambda^{(2;0)}$
given by the AC semiclassical theory in the DC limit can be simplified as
\begin{align}
&
\lim_{\omega_i \rightarrow 0}
J_\lambda^{(2;0)}
\nonumber \\
&=
\sum_n \int_k f_n
\left[
-\dfrac{\partial_\lambda a_n^{(1;\beta)}}{2}
+
\left(
\partial_\lambda a_n^{(1;\beta)}
-
\partial_\beta a_n^{(1;\lambda)}
\right)
\right]
E_\beta
\nonumber \\
&=
\sum_n \int_k
\left[
a_n^{(1;\lambda)}
\partial_\beta f_n
-
\dfrac{1}{2}
a_n^{(1;\beta)} \partial_\lambda f_n
\right]
E_\beta
\nonumber \\
&=
\sum_{nm} \int_k
\left(
\dfrac{g^{\lambda\alpha}_{nm}}{\epsilon_{nm}}\partial_\beta f_n
-
\dfrac{g^{\beta\alpha}_{nm}}{2\epsilon_{nm}}\partial_\lambda f_n
\right)
E_\alpha E_\beta
\nonumber \\
&=
\dfrac{1}{2}
\sum_{nm} \int_k
\left(
\dfrac{g^{\lambda\alpha}_{nm}}{\epsilon_{nm}}\partial_\beta f_n
+
\dfrac{g^{\lambda\beta}_{nm}}{\epsilon_{nm}}\partial_\alpha f_n
-
\dfrac{g^{\beta\alpha}_{nm}}{\epsilon_{nm}}\partial_\lambda f_n
\right)
E_\alpha E_\beta,
\label{J20DC}
\end{align}
where we have symmetrized the final result about the field indices.
Note that in the DC limit $a_n^{(1;\lambda)}=\sum_m g^{\lambda\alpha}_{nm}E_\alpha/\epsilon_{nm}$
with $g^{\lambda\alpha}_{nm}=r^\lambda_{nm}r^\alpha_{mn}+r^\alpha_{nm}r^\lambda_{mn}$ the quantum metric,
which is symmetric about $\lambda$ and $\alpha$.

On the other hand, by interchanging $m$ and $n$ and symmetrizing $\alpha$ and $\beta$, we find that
Eq.(\ref{sigmaeiint0}) can be expressed as
\begin{align}
\sigma^{(ei;int)}_{\lambda\alpha\beta}
&=
\sum_{mn} \int_k
\dfrac{1}{2}
\left(
\dfrac{g^{\lambda\beta}_{nm}}{\epsilon_{nm}} \partial_\alpha f_n
+
\dfrac{g^{\lambda\alpha}_{nm}}{\epsilon_{nm}} \partial_\beta f_n
\right).
\label{sigmaeiintDC}
\end{align}
In addition, for $\sigma^{(2;ie)}_{\lambda\alpha\beta}$,
in the DC limit we find
\begin{align}
\lim_{\omega_i \rightarrow 0}
\sigma^{(2;ie)}_{\lambda\alpha\beta}
=
\sum_{mn} \int_k \dfrac{f_{nm} r^\alpha_{mn}}{\epsilon_{nm}} \mathcal{D}^\beta_{nm}r^\lambda_{nm}.
\label{sigmaieDC}
\end{align}
Furthermore, for $\sigma^{(2;ee)}_{\lambda\alpha\beta}$, when $m=n$ we find
(see Appendix \ref{sigmaeemeqn})
\begin{align}
&
\sigma^{(2;ee)}_{\lambda\alpha\beta}(m=n)
\equiv \sigma^{(2;ee1)}_{\lambda\alpha\beta}
\nonumber \\
&=
\dfrac{1}{2}
\sum_{nm}
\int_k
\dfrac{\Delta^\lambda_{nm} f_{nm}r^\beta_{nm}r^\alpha_{mn}}
{\left( \omega_1-\epsilon_{mn} + i\xi \right)\left( \omega_2+\epsilon_{mn} + i\xi \right)},
\label{sigmaee1}
\end{align}
where $\Delta^\lambda_{nm}=v^\lambda_n-v^\lambda_m$.
As shown in Appendix \ref{sigmaeemeqn},
the divergent factor $1/(\omega_\Sigma+i2\xi)$ in $\sigma^{(2;ee)}_{\lambda\alpha\beta}(m=n)$
has been removed after symmetrization. In the DC limit, we then obtain
\begin{align}
\lim_{\omega_i \rightarrow 0} \sigma_{\lambda\alpha\beta}^{(2;ee1)}
&
\equiv
\dfrac{1}{2}
\sum_{nm}
\int_k
\dfrac{-\Delta^\lambda_{nm} f_{nm}r^\beta_{nm}r^\alpha_{mn}}{\epsilon_{nm}^2}
\nonumber \\
&=
\dfrac{1}{2}
\sum_{nm}
\int_k f_n g^{\beta\alpha}_{nm}
\partial_\lambda
\left(
\dfrac{1}{\epsilon_{nm}}
\right),
\label{sigmaeeint1DC}
\end{align}
which is symmetric about the field indices $\alpha$ and $\beta$.
When $m \neq n$, we can directly take the DC limit of $\sigma^{(2;ee)}_{\lambda\alpha\beta}$
and obtain
\begin{align}
&
\lim_{\omega_i\rightarrow 0}
\sigma_{\lambda\alpha\beta}^{(2;ee)}
(m \neq n)
\equiv
\lim_{\omega_i\rightarrow 0}
\sigma_{\lambda\alpha\beta}^{(2;ee2)}
\\
&=
\sum_{nml} \int_k ir^\lambda_{nm}
\left(
\dfrac{r^\beta_{ml} r^\alpha_{ln}f_{nl}}{-\epsilon_{ln}}
-
\dfrac{f_{lm}r^\alpha_{ml}r^\beta_{ln}}{-\epsilon_{ml}}
\right)
\nonumber \\
&=
\sum_{nml} \int_k
\dfrac{r^\alpha_{mn}f_{nm}}{\epsilon_{nm}}
i\left(
r^\lambda_{nl}r^\beta_{lm}
-
r^{\beta}_{nl}r^\lambda_{lm}
\right)
\nonumber \\
&=
\sum_{nm} \int_k
\dfrac{r^\alpha_{mn}f_{nm}}{\epsilon_{nm}}
\left(
\mathcal{D}^\lambda_{nm}r^\beta_{nm}
-
\mathcal{D}^\beta_{nm}r^\lambda_{nm}
\right),
\label{sigmaeeint2DC}
\end{align}
where we have used the relation \cite{Sipe2000}
\begin{align}
\mathcal{D}_{nm}^\lambda r^\beta_{nm} - \mathcal{D}_{nm}^\beta r^\lambda_{nm}
=
i\sum_l
\left(
r^\lambda_{nl} r^\beta_{lm}
-
r^\beta_{nl} r^\lambda_{lm}
\right).
\label{identity}
\end{align}
Combining Eq.(\ref{sigmaieDC}), Eq.(\ref{sigmaeeint1DC}) and Eq.(\ref{sigmaeeint2DC}), we find
\begin{align}
&
\lim_{\omega_i \rightarrow 0}
\left(
\sigma_{\lambda\alpha\beta}^{(2;ie)}
+
\sigma_{\lambda\alpha\beta}^{(2;ee)}
\equiv
\sigma_{\lambda\alpha\beta}^{(2;ie)}
+
\sigma_{\lambda\alpha\beta}^{(2;ee1)}
+
\sigma_{\lambda\alpha\beta}^{(2;ee2)}
\right)
\nonumber \\
&=
\sum_{nm} \int_k
\left[
\dfrac{r^\alpha_{mn}f_{nm}}{\epsilon_{nm}}
\mathcal{D}^\lambda_{nm}r^\beta_{nm}
+
\dfrac{1}{2}
f_n g^{\beta\alpha}_{nm}
\partial_\lambda
\left(
\dfrac{1}{\epsilon_{nm}}
\right)
\right]
\nonumber \\
&=
\dfrac{1}{2}
\sum_{nm}
\int_k f_n
\left[
\dfrac{1}{\epsilon_{nm}}
\partial_\lambda g^{\beta\alpha}_{nm}
+
 g^{\beta\alpha}_{nm}
\partial_\lambda
\left(
\dfrac{1}{\epsilon_{nm}}
\right)
\right]
\nonumber \\
&=
-\dfrac{1}{2}
\sum_{nm}
\int_k
\dfrac{g^{\beta\alpha}_{nm}}{\epsilon_{nm}}
\partial_\lambda f_n,
\label{sigmaeeieintDC}
\end{align}
where we have symmetrized the first term of the second line about $\alpha$ and $\beta$
to obtain the final result and used the following relation \cite{Sipe2000}
\begin{align}
\partial_\lambda(r^\alpha_{nm}r^\beta_{mn})
=
\left(\mathcal{D}^\lambda_{nm}r^\alpha_{nm}\right)r^\beta_{mn}
+
r^\alpha_{nm}\left(\mathcal{D}^\lambda_{mn}r^\beta_{mn}\right).
\end{align}
By further combining Eq.(\ref{sigmaeeieintDC}) with Eq.(\ref{sigmaeiintDC}),
we finally arrive tat
\begin{align}
&
\lim_{\omega_i \rightarrow 0}
\left[
\sigma_{\lambda\alpha\beta}^{(2;ie)}
+
\sigma_{\lambda\alpha\beta}^{(2;ee)}
+
\sigma_{\lambda\alpha\beta}^{(ei;int)}
\right]
E_\alpha^{\omega_1} E_\beta^{\omega_2}
\nonumber \\
&=
\dfrac{1}{2}
\sum_{nm} \int_k
\left(
\dfrac{g^{\lambda\alpha}_{nm}}{\epsilon_{nm}}\partial_\beta f_n
+
\dfrac{g^{\lambda\beta}_{nm}}{\epsilon_{nm}}\partial_\alpha f_n
-
\dfrac{g^{\beta\alpha}_{nm}}{\epsilon_{nm}}\partial_\lambda f_n
\right)
E_\alpha E_\beta
\nonumber \\
&=
\lim_{\omega_i \rightarrow 0} J^{(2;0)}_{\lambda},
\end{align}
which shows that the intrinsic second-order nonlinear current obtained with
the response theory is also equivalent to that derived with the AC semiclassical theory,
particularly in the DC limit.
Note that this intrinsic second-order nonlinear current
contains both the longitudinal (from the second-order energy correction $\epsilon_n^{(2)}$)
and transverse (from both the anomalous velocity $\vect{E}\times\vect{\Omega}_n^{(1)}$
and the second-order energy correction $\epsilon_n^{(2)}$) components,
which can only appear in systems without $\mathcal{P}$ ($\mathcal{P}$, inversion) and $\mathcal{T}$ symmetries.

At this stage, we conclude that the intrinsic second-order nonlinear longitudinal current
proposed in Refs. \cite{CulcerPRBL, GangSu2022} based on the response theory
can also be reproduced by the semiclassical theory when the second-order wavepacket energy is
taken into account (ignored in Ref. \cite{CulcerPRBL}).
As a result, the response theory does not contain any interband coherent correction
compared to the semiclassical theory.
We stress that the intrinsic contributions given by both formulations are irrelevant to the relaxation
so that there is no need for regulating ($\xi \rightarrow 0$ does not lead to a divergence)
even in the DC limit.
In other words, the semiclassical theory give the same intrinsic physics as the response theory.
Finally, to clearly illustrate the equivalence between two formulations,
we schematically summarized the main results obtained in this subsection into FIG.(\ref{FIG1}),
which will be useful when we show that both formulations are equivalent
in the AC regime in the following.

\subsection{The second order: the AC case}

Guided by the equivalence between the semiclassical theory
and the response theory in the DC limit, as summarized in FIG.(\ref{FIG1}),
we further show the equivalence between both approaches in the AC regime.
First of all, it is easy to show that
\begin{align}
\sigma_{\lambda\alpha\beta}^{(2;ii)} E_\alpha^{\omega_1} E_\beta^{\omega_2}
=
\lim_{\eta \rightarrow 0} J_{\lambda}^{(2;2)}
\end{align}
by using Eq.(\ref{sigmaii}) and Eq.(\ref{J22}), in which Eq.(\ref{neq2}) is inserted.

In the DC case, we have shown that the conductivity $\sigma^{(2;ei)}_{\lambda\alpha\beta}$
can be divided into the extrinsic and intrinsic contributions.
In the AC case, we follow the same partition. In addition,
we may expect some additional contributions that vanish in the DC limit,
which should correspond to the displacement current given by the time variation of
\textit{positional shift} in the AC semiclassical theory.
Particularly, using Eq.(\ref{factorization}) and
\begin{align}
\dfrac{\epsilon_{mn}}{\omega_2-\epsilon_{mn}+i\xi}
=
-1
+
\dfrac{\omega_2+i\xi}{\omega_2-\epsilon_{mn}+i\xi},
\end{align}
we can partition $\sigma_{\lambda\alpha\beta}^{(2;ei)}$ into
$
\sigma^{(2;ei)}_{\lambda\alpha\beta}
\equiv
\sigma^{(ei;1)}_{\lambda\alpha\beta}
+
\sigma^{(ei;0)}_{\lambda\alpha\beta}
+
\sigma^{(ei;\bar{1})}_{\lambda\alpha\beta}
$,
where
\begin{align}
\sigma^{(ei;1)}_{\lambda\alpha\beta}
&=
\sum_{mn} \int_k
\dfrac{-r^\lambda_{nm}r^\beta_{mn} \partial_\alpha f_{nm}}
{\omega_1+i\xi},
\\
\sigma^{(ei;\bar{1})}_{\lambda\alpha\beta}
&=
\sum_{mn} \int_k
\dfrac{(\omega_2+i\xi)r^\lambda_{nm}r^\beta_{mn}\partial_\alpha f_{nm}}
{(\omega_2-\epsilon_{mn}+i\xi)\left(\omega_1+i\xi\right)},
\\
\sigma^{(ei;0)}_{\lambda\alpha\beta}
&\equiv
\sum_{mn} \int_k
\dfrac{-\epsilon_{mn}r^\lambda_{nm}r^\beta_{mn}\partial_\alpha f_{nm}}
{(\omega_2-\epsilon_{mn}+i\xi)\left(\omega_\Sigma-\epsilon_{mn}+i2\xi\right)}
\nonumber \\
&=
\sum_{mn} \int_k
\dfrac{-iv^\lambda_{nm}r^\alpha_{mn}\partial_\beta f_{nm}}
{(\omega_1-\epsilon_{mn}+i\xi)\left(\omega_\Sigma-\epsilon_{mn}+i2\xi\right)}
\label{sigmaei0}.
\end{align}
Here the superscripts $(ei;1)$, $(ei;\bar{1})$, and $(ei;0)$ stand for
the extrinsic (existing in the DC limit), AC (disappearing in the DC limit), and intrinsic (existing in the DC limit)
contributions, respectively. As expected, we find
\begin{align}
\left(
\sigma_{\lambda\alpha\beta}^{(ei;1)}
+
\sigma_{\lambda\alpha\beta}^{(ei;\bar{1})}
\right)
E_\alpha^{\omega_1}
E_\beta^{\omega_2}
=
\lim_{\eta \rightarrow 0} J_\lambda^{(2;1)},
\end{align}
where
\begin{align}
&
\lim_{\eta \rightarrow 0} J_\lambda^{(2;1)}
=
\lim_{\eta \rightarrow 0}
\sum_{n} \int_k f_n^{(1)}
\left(
\Omega^{\lambda\beta}_n E_\beta^{\omega_2}
+
\partial_t a_n^{(1;\lambda)}
\right)
\nonumber \\
&=
\sum_{n} \int_k
\dfrac{-r^\lambda_{nm}r^\beta_{mn} \partial_\alpha f_{nm}}{\omega_1+i\xi}
E_\alpha^{\omega_1}
E_\beta^{\omega_2}
\nonumber \\
&+
\sum_{mn} \int_k
\dfrac{(\omega_2+i\xi)r^\lambda_{nm}r^\beta_{mn}\partial_\alpha f_{nm}}
{( \omega_2 - \epsilon_{mn} + i\xi )(\omega_1+i\xi)}
E_\alpha^{\omega_1}
E_\beta^{\omega_2}
\end{align}
by inserting Eq.(\ref{an1}) and Eq.(\ref{neq1}) into Eq.(\ref{J21}).
Note that $\Omega^{\lambda\beta}_n=\sum_m i(r^\lambda_{nm}r^\beta_{mn}-r^{\beta}_{nm}r^{\lambda}_{mn})$
and the second line is obtained by index interchanging.

For the remaining terms both in the response theory
and in the AC semiclassical theory, we can also establish
the following equivalent relation (see Appendix \ref{intrinsicAC})
\begin{align}
\left(
\sigma_{\lambda\alpha\beta}^{(2;ee)}
+
\sigma_{\lambda\alpha\beta}^{(2;ie)}
+
\sigma_{\lambda\alpha\beta}^{(ei;0)}
\right)
E_\alpha^{\omega_1}
E_\beta^{\omega_2}
=
J_\lambda^{(2;0)},
\label{J20eeeiie}
\end{align}
which is the same as the DC case, as illustrated in FIG.(\ref{FIG1}).

As expected, the AC semiclassical theory is also equivalent to
the response theory at the second order of the electric field,
particularly in the clean limit $\eta \rightarrow 0$.
Beyond this limit, we can still introduce the relaxation in the response theory
in the same way as we have adopted in the DC limit.
By doing that, the semiclassical results at both the zero and finite frequencies can be
reproduced by the response theory.
Besides this manual way based on the equivalence between both formulations,
we wish to remark that the previous relaxation schemes adopted in the response theory 
particularly by modifying the quantum Liouville equation in fact is unresonable,
as will be explained in section \ref{relaxationres}.

\bigskip
\section{Consequences of the equivalence}
\label{consequence}

In this section, we discuss the implications behind the equivalence
between the AC semiclassical theory and the
response theory. First, we show that the intrinsic nonlinear longitudinal current
and the nonreciprocal magnetoresistance represent the same physics.
Second, we suggest that the energy used in the equilibrium Fermi distribution
in the semiclassical theory is the unperturbed band energy
by assuming that both formulations give the same intrinsic responses.
Third, we argue that phenomenally modifying the quantum Liouville equation
to introduce relaxation in the response theory needs to be reconsidered.

\subsection{Unifying the intrinsic nonlinear longitudinal current}
\label{Kohn}

In the previous section, we have shown that the intrinsic nonlinear longitudinal current,
as first derived with the response theory \cite{CulcerPRBL, GangSu2022},
can be obtained by the semiclassical theory with the consideration of
the second-order wavepacket energy correction,
which has been suppressed in the pioneering paper Ref. \cite{GaoY2014} 
when evaluating the intrinsic second-order nonlinear current.
Interestingly, rather than using the wavepacket method,
we note that the second-order energy calculated with the Luttinger-Kohn method \cite{YBH2ndenergy},
can also lead to an intrinsic nonlinear longitudinal current (nonreciprocal magnetoresistance \cite{YBH2ndenergy}), 
which has been added to the extended second-order semiclassical results given by Ref. \cite{GaoY2014}
due to the same quantum metric origin.
And this second-order energy correction also plays a key role
in understanding the gravitational anomaly \cite{ganomaly}.

Unfortunately, we wish to point out that the energy calculated with the Luttinger-Kohn method
can not be directly used in the semiclassical theory. The wavepacket energy in the semiclassical theory
accurate up to the second order of the electric field is given by \cite{XiaoC2ndenergy, XiaoCnormal}
\begin{align}
\bar{\epsilon}_n
=
\epsilon_n + \vect{E} \cdot \vect{r}_c
-
\dfrac{1}{2} \sum_{m} \dfrac{g_{nm}^{\alpha\beta}E_\alpha E_\beta}{\epsilon_{nm}},
\end{align}
where we take the DC limit and
\begin{align}
\vect{r}_c = \partial_{\vect{k}_c} \gamma + \vect{\mathcal{A}}_n + \vect{a}_n^{(1)} + \vect{a}_n^{(2)} \cdots. 
\end{align}
As stated previously, $\vect{E}\cdot\vect{r}_c$ does not enter the semiclassical equations of motion
since $\vect{r}_c$ and $\vect{k}_c$ are independent variables in the Lagrangian \cite{Xiao2010}.
Explicitly, the energy that appeared in the semiclassical equations of motion is \cite{XiaoC2ndenergy, XiaoCnormal}
\begin{align}
\bar{\epsilon}_n^{\text{EOM}}
=
\epsilon_n 
-
\dfrac{1}{2} \sum_{m} \dfrac{g_{nm}^{\alpha\beta}E_\alpha E_\beta}{\epsilon_{nm}}.
\end{align}
On the other hand, the energy calculated with the Luttinger-Kohn method is given by
\begin{align}
\bar{\epsilon}_n^{\text{KL}}
=
\epsilon_n 
+
\dfrac{1}{2} \sum_{m} \dfrac{g_{nm}^{\alpha\beta}E_\alpha E_\beta}{\epsilon_{nm}},
\label{KLenergy}
\end{align}
which is reproduced in Appendix \ref{Kohnenergy} by exactly following Ref. \cite{YBH2ndenergy}.
Compared to the semiclassical wavepacket energy $\bar{\epsilon}^{\text{EOM}}_n$,
we note that the second-order energy in $\bar{\epsilon}_{n}^{\text{KL}}$ differs a sign
with that in $\bar{\epsilon}_n^{\text{EOM}}$.
Notably, we note that $\bar{\epsilon}_n^{\text{KL}}$ 
with the Luttinger-Kohn method bypasses the gauge-variant energy correction
$\vect{E}\cdot\vect{\mathcal{A}}_n$ by choosing a special gauge.
And it is this gauge choice that leads to the inconsistency
between $\bar{\epsilon}_n^{\text{EOM}}$ and $\bar{\epsilon}_n^{\text{KL}}$
since the gauge-variant energy $\vect{E}\cdot\vect{r}_c$ bypassed in the semiclassical theory
also contains a second-order correction
\begin{align}
\Delta \bar{\epsilon}_n^{(2)} 
=
\vect{E} \cdot \vect{a}_n^{(1)} = \sum_n \dfrac{g^{\alpha\beta}_{nm}E_\alpha E_\beta}{\epsilon_{nm}}.
\end{align}
By removing this correction from $\bar{\epsilon}_n^{\text{KL}}$,
we find \cite{xiang}:
\begin{align}
\bar{\epsilon}_n^{\text{KL}}
-
\Delta \bar{\epsilon}_n^{(2)} 
=
\epsilon_n 
-
\dfrac{1}{2} \sum_{m} \dfrac{g_{nm}^{\alpha\beta}E_\alpha E_\beta}{\epsilon_{nm}}
=
\bar{\epsilon}_n^{\text{EOM}}.
\end{align}
In section \ref{equivalence}, we have shown that the second-order wavepacket energy
in the semiclassical theory is responsible for the intrinsic nonlinear longitudinal current,
as firstly derived by the response theory. As a result, the nonreciprocal magnetoresistance
(intrinsic nonlinear longitudinal current)
that arises from the same energy correction represents the same physics with
the intrinsic nonlinear longitudinal current.
Particularly, by using Eq.(\ref{J20DC}) and
defining $\lim_{\omega_i \rightarrow 0} J_\lambda^{(2;0)} \equiv \sigma^\lambda_{\alpha\beta}E_\alpha E_\beta$,
we find
\begin{align*}
\sigma_{\alpha\beta}^\lambda
&=
\dfrac{1}{2}
\sum_{nm} \int_k
\left(
\dfrac{g^{\lambda\alpha}_{nm}}{\epsilon_{nm}}\partial_\beta f_n
+
\dfrac{g^{\lambda\beta}_{nm}}{\epsilon_{nm}}\partial_\alpha f_n
-
\dfrac{g^{\beta\alpha}_{nm}}{\epsilon_{nm}}\partial_\lambda f_n
\right),
\end{align*}
which contains both the intrinsic nonlinear Hall and longitudinal (or nonreciprocal magnetoresistance) currents.

\subsection{Energy used in the distribution function}

Note that in the previous sections, we have assumed that the energy used
in the equilibrium Fermi distribution function is the unperturbed band energy.
Recently, some works obtained the third-order nonlinear Hall effects \cite{BPT, TaIrTe4, Liu2022, Nandy}
by Taylor expanding the Fermi distribution function where the energy argument of
the equilibrium Fermi distribution function takes into account
the correction due to the external field.
For instance, by replacing the unperturbed energy in $f_n^{(1)}$ with $\bar{\epsilon}_n=\epsilon_n+\epsilon_n^{(2)}$
and further Taylor expanding $f_n^{(1)}(\bar{\epsilon}_n)$ as
\begin{align}
f_n^{(1)} (\bar{\epsilon}_n)
&=
\tau E_\alpha \partial_\alpha f_n (\bar{\epsilon}_n)
\nonumber \\
&=
\tau E_\alpha \partial_\alpha f_n (\epsilon_n)
+
\tau E_\alpha \partial_\alpha \left[ \epsilon_n^{(2)}f_n' (\epsilon_n) \right],
\label{feq1exp}
\end{align}
one can immediately obtain an extrinsic third-order nonlinear current
\begin{align}
J_\lambda = \tau \sum_n \int_k  v_n^\lambda E_\alpha \partial_\alpha \left[ \epsilon_n^{(2)}f_n' (\epsilon_n) \right].
\end{align}
Note that Eq.(\ref{feq1exp}) taken the DC limit of Eq.(\ref{neq1}). 
However, based on the equivalence between both formulations,
we argue that these third-order extrinsic current responses can not appear. 

In particular, if the energy argument used in the equilibrium Fermi distribution
function considers the correction under the electric field,
it is easy to show that \cite{GaoY2014, XiaoCPRLPHE}:
\begin{align}
\int_k f_n(\bar{\epsilon}_n) \bar{\vect{v}}_n^a
=
\int_k f_n(\bar{\epsilon}_n) \dfrac{\partial \bar{\epsilon}_n}{\partial \vect{k}}
=
\int_k \dfrac{\partial g_n(\bar{\epsilon}_n)}{\partial \vect{k}}
=
0,
\label{suppress}
\end{align}
where $g_n=-k_BT \ln \left[ 1 + e^{-(\bar{\epsilon}_n-\mu)/k_BT} \right]$ \cite{XiaoCnormal} and
the band summation over $n$ is suppressed for brevity.
Clearly, Eq.(\ref{suppress}) forbids the intrinsic nonlinear longitudinal current
arising from the field-induced group velocity,
but these intrinsic nonlinear longitudinal currents can be allowed by the response theory.
As a result, by assuming that both formulations give the same intrinsic physics,
we suggest that the energy used in the equilibrium Fermi distribution in the semiclassical theory
is the unperturbed band energy, as we have adopted in section \ref{currentdensity} and section \ref{equivalence}.

\subsection{Relaxation in the response theory}
\label{relaxationres}

Finally, we discuss the phenomenological relaxation introduced in the response theory.
We mention that the conductivity tensors Eqs.(\ref{sigmaee}-\ref{sigmaii}) from the response theory
are obtained by solving the quantum Liouville equation Eq.(\ref{QL0})
and hence do not contain relaxation like the semiclassical theory,
where the relaxation is explicitly introduced through solving the Boltzmann equation.
Usually, to take the relaxation into account in the response theory,
Eq.(\ref{QL0}) is phenomenologically modified as
\cite{Mikhailov, Peres}
\begin{align}
i\partial_t \rho_{mn} = [H, \rho]_{mn} -\dfrac{i}{\tau} (\rho_{mn} - \rho^{(0)}_{mn}),
\label{modify1}
\end{align}
which has been used in many studies \cite{Sodemann2019, HWXu2021, Culcer2022} to evaluate the nonlinear responses.
Interestingly, we find that this phenomenological modification
will give (see Appendix \ref{modifyapp}),
\begin{align}
\sigma_{\lambda\alpha}^H
&
=
2 \sum_{mn} \int_k f_n \text{Re} \left[\dfrac{r^\alpha_{mn}v^\lambda_{nm}}{-\epsilon_{mn}+i/\tau}\right]
\label{L-Hall}.
\end{align}
We have the following observations about this result. 
First, Eq. (\ref{L-Hall}) can deliver a linear extrinsic Hall current in $\mathcal{T}$-invariant systems
(Note that $\sigma_{\lambda\alpha}^H$ is a $\mathcal{T}$-even rank-2 tensor),
which contradicts the current experimental recognition and is also not
allowed by the Onsager reciprocal relation \cite{Nagaosa2010, XiaoCnc, Onsager}.
Second, Eq. (\ref{L-Hall}) can give a $\tau^{-1}$-dependent current responses
when $\hbar/\tau \gg \epsilon_{mn}$, 
which can not show up in the semiclassical results but surprisingly
appear in the nonlinear current responses \cite{Yanase2020} based on the response theory.
Third, the equivalence between both formulations is broken.
Therefore, we argue that the additional term in Eq.(\ref{modify1}) is unreasonable.

Interestingly, if the phenomenological relaxation introduced in Eq.(\ref{modify1})
is constrained to the intraband scattering, namely by modifying Eq.(\ref{modify1}) into
\begin{align}
i\partial_t \rho_{mn} = [H, \rho]_{mn} - \frac{i}{\tau} \delta_{mn} (\rho_{mn} - \rho^{(0)}_{mn}),
\label{modify2}
\end{align}
we find that the response theory
agrees perfectly with the semiclassical theory in the linear order of the electric field,
as shown in Appendix \ref{modifyapp}.
Furthermore, at the second order of the electric field, Eq.(\ref{modify2}) gives (see Appendix \ref{modifyapp}),
\begin{align}
\sigma_{\lambda\alpha\beta}^{(2,ee)}
&=
\sum_{mnl}
\int_k
\bar{\mathcal{V}}_{nm}^\lambda
\left(\dfrac{r^\beta_{ml} r^\alpha_{ln}f_{nl} }{\omega_1-\epsilon_{ln}+i\xi}
-
\dfrac{f_{lm}r^\alpha_{ml}r^\beta_{ln}}{\omega_1 -\epsilon_{ml}+i\xi}
\right),
\label{sigmaeemodify2}
\\
\sigma_{\lambda\alpha\beta}^{(2,ei)}
&=
\sum_{mn}
\int_k
\bar{\mathcal{V}}_{nm}^\lambda
\left(\dfrac{ir^\beta_{mn}\partial_{\alpha}f_{nm}}{\omega_1+i(\eta+\xi)}\right),
\label{sigmaeimodify2}
\\
\sigma_{\lambda\alpha\beta}^{(2,ie)}
&=
\sum_{mn} \int_k
\bar{\mathcal{V}}_{nm}^\lambda
\mathcal{D}^{mn}_\beta
\left(
\dfrac{if_{nm}r^\alpha_{mn}}{\omega_1-\epsilon_{mn}+i\xi}
\right),
\label{sigmaiemodify2}
\\
\sigma_{\lambda\alpha\beta}^{(2,ii)}
&=
\sum_{n} \int_k
\dfrac{-v^\lambda_{n}\partial^2_{\alpha \beta}f_n}{\left[\omega_1+i(\eta+\xi)\right]\left[\omega_\Sigma+i(\eta+2\xi)\right]},
\label{sigmaiimodify2}
\end{align}
where $\bar{\eta}=\eta\delta_{mn}$ (note that $\eta=1/\tau$) and
\begin{align}
\bar{\mathcal{V}}_{nm}^\lambda
\equiv
\dfrac{v^\lambda_{nm} }{\omega_\Sigma-\epsilon_{mn}+i(\bar{\eta}+2\xi)}.
\end{align}
At this stage, we first note that Eqs. (\ref{sigmaiimodify2}) and (\ref{sigmaeimodify2}), respectively,
can also perfectly reproduce the extrinsic nonlinear Drude and Hall currents given by
Eqs. (\ref{J22}) and (\ref{J21}) in the AC semiclassical theory, as outlined in FIG. \ref{FIG1}.
Note that that Eq. (\ref{sigmaiimodify2}) also contains the intrinsic contribution, see FIG. \ref{FIG1}.
Unfortunately, we find that Eq. (\ref{sigmaeemodify2}) particularly with $m=n$
can no longer give an intrinsic current like Eq. (\ref{sigmaeeint1DC})
given by the original quantum Liouville equation, although the intrinsic currents
contributed by Eqs. (\ref{sigmaeimodify2}) and (\ref{sigmaiemodify2}) keep unchanged.
Therefore, the intrinsic current given by the semiclassical theory
can no longer be reproduced by the modified response theory,
particularly in terms of Eq. (\ref{modify2}).

In fact, in section \ref{equivalence}, a scheme to include relaxation in the response theory
has been utilized,
particularly by regulating the clean zero-frequency divergent conductivities
given by both formulations on the same footing.
However, we can not rule out other ways of introducing relaxation as long as
it will not lead to unphysical results. To close this section,
we believe that the distinction between the semiclassical theory
and the response theory is rooted in the different ways of including relaxation.
We conclude that the previous scheme to include relaxation in the response theory
by modifying the Liouville equation for the nonequilibrium density matrix needs to be reconsidered.

\bigskip
\section{AC nonlinear effects}
\label{ACcorrection}

After establishing the equivalence between both formulations, in this section,
we focus on the overlooked second-order nonlinear current responses unique to the AC transport,
which contains the intrinsic second-order nonlinear displacement current from the time variation of
the second-order \textit{positional shift} and the extrinsic second-order nonlinear displacement
current, which is obtained by combining the first-order nonequilibrium Fermi distribution with
the time variation of the first-order \textit{positional shift}. Although both formulations
can give these two contributions, we will use the expressions given by the AC semiclassical theory
since it gives a clear partition and presents a clear physical origin of the \textit{positional shift}.

\subsection{The intrinsic nonlinear displacement current}

From the extended AC semiclassical theory, the intrinsic nonlinear displacement current is given by
\begin{align}
J_\lambda^{(2;\bar{0})} = \sum_n \int_k f_n \partial_t a_n^{(2;\lambda)} 
=-\omega \sigma_{\alpha\beta}^{\lambda} E_\alpha E_\beta^\omega,
\end{align}
where $E_\alpha$ and $E_\beta^\omega = E_\beta \sin(\omega t)$ stands for the applied DC and AC electric fields, respectively.
Under this crossed setup, we find
\begin{align}
\sigma_{\alpha\beta}^{\lambda}
&=
\sum_{nml}\int_k f_n
\text{Re}
\left(
\dfrac{2r^\lambda_{nm}r^\beta_{ml}r^\alpha_{ln}+r^\beta_{nl} r^\lambda_{lm} r^\alpha_{mn}}{\epsilon_{nl}\epsilon_{nm}}
\right)
\nonumber \\
&+
\sum_{nm}\int_k f_n
\text{Re}
\left[
i\mathcal{D}^\beta_{mn}
\left(
\dfrac{r^\alpha_{mn}}{\epsilon_{nm}}
\right)
\dfrac{2r^\lambda_{nm}}{\epsilon_{nm}}
\right]
\nonumber \\
&-
\sum_{nm}\int_k f_n
\text{Re}
\left[
i\mathcal{D}^\lambda_{nm}
\left(
\dfrac{r^\alpha_{nm}}{\epsilon_{nm}}
\right)
\dfrac{r^\beta_{mn}}{\epsilon_{nm}}
\right],
\end{align}
which is obtained by taking $\omega_1=0$ and summing over $\omega_2=\pm \omega$ in Eq. (\ref{an2}).
In addition, $\hbar \omega \ll \epsilon_{mn}$ is assumed.

The same as the linear intrinsic
displacement current \cite{XiangDHE}, we note that $\sigma^{\gamma}_{\alpha\beta}$ also features the
$\mathcal{T}$-even ($\mathcal{T}$, time reversal) behavior,
which indicates that this nonlinear current can appear in $\mathcal{T}$-invariant systems but
without $\mathcal{P}$-symmetry. Importantly, by using the Neumann's principle,
we note that $\sigma_{\alpha\beta}^{\gamma}$ can deliver an intrinsic nonlinear Hall current
in $\mathcal{T}$-invariant systems with the crystalline point groups of $1$, $2$, $m$, $222$, $mm2$,
$4$, $-4$, $422$, $4mm$, $-42m$, $3$, $32$, $3m$, $6$, $-6$, $622$, $6mm$, $-6m2$, $23$, $432$, $-43m$.
Furthermore, this intrinsic nonlinear displacement current
can coexist with the $\mathcal{T}$-even
Berry-curvature-dipole driven extrinsic nonlinear Hall current \cite{BCD}
particularly under an AC electric field.

\subsection{The extrinsic nonlinear displacement current}

In addition to the intrinsic nonlinear displacement current,
the AC semiclassical theory also delivers an extrinsic nonlinear displacement current,
which is given by
\begin{align}
J_\lambda^{(2;\bar{1})}
& \equiv
\sum_n \int_k f_n^{(1)} \partial_t a_n^{(1;\lambda)}
=-\omega \tau \bar{\sigma}_{\alpha\beta}^{\gamma} E_\alpha E_\beta^\omega,
\end{align}
where we have taken $\omega_1=0$ in Eq.(\ref{neq1}) and summed over $\omega_2=\pm \omega$ appeared in $a_n^{(1;\lambda)}$.
By doing this, we find
\begin{align}
\bar{\sigma}_{\alpha\beta}^{\gamma}
=
\sum_{nm}
\int_k
\partial_\alpha f_{n} 
\dfrac{g^{\beta\lambda}_{nm}}
{\epsilon_{nm}}
\label{nonlineardispext}
\end{align}
by further assuming $\hbar \omega \ll \epsilon_{mn}$.

Interestingly, we find that Eq. (\ref{nonlineardispext}) is also solely driven by the quantum metric dipole
like the intrinsic nonlinear longitudinal and Hall currents.
As a result, the corresponding conductivity tensors
exactly respect the same symmetry transformation under the magnetic point group operations.
However, we note that the nonlinear extrinsic displacement current scales with a dimensionless constant $\omega \tau$,
which means that the extrinsic nonlinear current will be significantly enhanced when $\omega \tau \gg 1$
under an AC electric field while the intrinsic nonlinear longitudinal and Hall currents will not be affected.

\section{Discussion and summary}
\label{summary}

First of all, although we stop at the second order of the electric field, we believe
that the equivalence between both formulations can be extended to a higher order,
such as the third order.
Second, we wish to mention that the response theory used in this work
is formulated at the length gauge.
Recently, some second-order nonlinear currents \cite{Yanunify, Kapla2023}
based on the velocity-gauge response theory are proposed,
which can also not be reproduced by the AC semiclassical theory
and are treated as the quantum correction beyond the semiclassical theory.
However, based on the equivalence between both formulations,
we believe that these quantum corrections may also be rooted in different ways of introducing phenomenological relaxation.
In addition, we wish to remark that the $\tau^{-1}$-dependent nonlinear current
obtained with response theory at the length gauge may also be related
to the different schemes of introducing relaxation \cite{Yanase2020}.
Finally, we did not discuss the consequences of the side jump and skew scattering
\cite{disorderCulcer3,Nagaosa2010, JTsong2023, LuHZNC, XiaoCdisorder}
which is beyond the scope of this work.
In addition, the equivalence between both formulations with the consideration of magnetic field \cite{Bequ}
particularly at nonlinear regimes may be explored in the future.
Finally, a detailed study of the intrinsic and extrinsic nonlinear displacement currents
also will be investigated in the future.

In summary, by extending the linear AC semiclassical theory under a uniform electric field
to the nonlinear regime, we show that up to the second order of the electric field,
the AC semiclassical theory is equivalent to the response theory in the clean limit and
this equivalence can be expected even in the DC limit when the zero-frequency divergence is properly
regulated by introducing the relaxation on the same footing.
In particular, in the linear order of the electric field, we show that both formulations
can give the linear Drude current, the intrinsic anomalous Hall current, and the intrinsic displacement current,
in which the last one arising from the time variation of the first-order \textit{positional shift}
can only be driven by an AC field and is usually overlooked even in AC transport.
Furthermore, in the second order of the electric field, we show that all the DC second-order nonlinear current responses,
including the nonlinear Drude current, the nonlinear Hall currents that
driven by the Berry curvature dipole and quantum metric dipole, respectively,
and the nonlinear intrinsic longitudinal current response,
can be obtained by both formulations.
Furthermore, we show that both formulations can
give the extrinsic and intrinsic  nonlinear displacement currents,
both of which can only appear in AC transport and are not reported yet.
Notably, we find that the extrinsic nonlinear displacement current is also
solely driven by the quantum metric dipole but scales with a dimensionless constant $\omega\tau$.
As a result, both the quantum-metric-dipole driven
intrinsic nonlinear Hall and longitudinal currents under an AC electric field
can be significantly enhanced by this extrinsic
nonlinear displacement current when $\omega \tau \gg 1$.
Besides, we note that the $\mathcal{T}$-even
intrinsic nonlinear displacement current
can coexist with the Berry-curvature-dipole driven extrinsic nonlinear Hall current
under an AC electric field. 

On top of this equivalence,
(i) we suggest that the energy used in the equilibrium Fermi distribution
in the semiclassical theory is the unperturbed band energy
by assuming that both formulations give the same intrinsic responses; 
(ii) we unify the intrinsic second-order nonlinear longitudinal current responses calculated with different approaches; 
(iii) we argue that the scheme of introducing relaxation in the response theory
by modifying the Liouville equation for the nonequilibrium density matrix needs to be reconsidered.
Our work clarifies the conceptual puzzles encountered
in the emergent field of \textit{nonlinear Hall effects} and highlights that
the displacement nonlinear current responses unique to the AC transport
can be important especially when a high-frequency AC driving field is applied.

\newpage
\onecolumngrid

\section{Appendix}
\label{appendix}

\appendix

\section{Calculation details in Section \ref{ACsemitheory}}

\subsection{Eqs.(\ref{rcdef}-\ref{an2})}
\label{pcenter}

With the normalized second-order wavepacket Eq.(\ref{wavepacket2}),
the position center $\vect{r}_c$ can be expressed as
\begin{align}
\vect{r}_c
\equiv
\langle \bar{W}|\vect{r}|\bar{W}\rangle
=
(1+2\delta)\langle W_0|\vect{r}|W_0\rangle
+
2\text{Re}\langle W_0|\vect{r}|W_1\rangle
+
\langle W_1|\vect{r}|W_1\rangle
+
2\text{Re}\langle W_0|\vect{r}|W_2\rangle
+
\mathcal{O}(E_\alpha^3).
\end{align}
For $\langle W_0|\vect{r}|W_0\rangle$, we find \cite{Xiao2010}
\begin{align}
\langle W_0|\vect{r}|W_0\rangle
&=
\int_{\vect{p}}\int_{\vect{p}'} C_{n}(\vect{p}) C_n^*(\vect{p}') e^{i(\vect{p}-\vect{p}')\cdot\vect{r}}
\langle u_n'|\vect{r}|u_n\rangle
=
\int_{\vect{p}}\int_{\vect{p}'} C_{n}(\vect{p}) C_n^*(\vect{p}')
\left[-i\partial_{\vect{p}} + \vect{\mathcal{A}}_n \right]
\delta(\vect{p}-\vect{p}')
\nonumber \\
&=
\partial_{\vect{p}_c}\gamma(\vect{p}_c)+\vect{\mathcal{A}}_n(\vect{p}_c),
\end{align}
where we have used $|C_n(\vect{p})|^2=\delta(\vect{p}-\vect{p}_c)$.
In addition, we have assumed that $C_n(\vect{p})=|C_n(\vect{p})|e^{-i\gamma(\vect{p})}$ and used that
\begin{align}
\int_{\vect{p}}\int_{\vect{p}'} C_{n}(\vect{p}) C_n^*(\vect{p}')
(-i\partial_{\vect{p}})\delta(\vect{p}-\vect{p}')
&=
\int_{\vect{p}} \left[ i\partial_{\vect{p}} C_n(\vect{p}) \right] C_n^*(\vect{p})
=
\int_{\vect{p}} \partial_{\vect{p}}\gamma (\vect{p}) |C_n(\vect{p})|^2
+
\int_{\vect{p}} |C_n(\vect{p})| i \partial_{\vect{p}}|C_n(\vect{p})|
\nonumber \\
&=
\partial_{\vect{p}_c}\gamma(\vect{p}_c)
+
\dfrac{i}{2} \partial_{\vect{p}} \left( \int_{\vect{p}} |C_n(\vect{p})|^2 \right)
=
\partial_{\vect{p}_c}\gamma(\vect{p}_c).
\end{align}
For $2\text{Re}\langle W_0|\vect{r}|W_1\rangle$, we find
\begin{align}
2\text{Re}\langle W_0|\vect{r}|W_1\rangle
=
2\text{Re}
\left[
\int_{\vect{p}} \int_{\vect{p}'}
\sum_{m \neq n}
M_{mn}^{(1)}(\vect{p})
C_n(\vect{p})C_n^*(\vect{p}')
e^{i(\vect{p}-\vect{p}') \cdot \vect{r}} \langle
u_n'|\vect{r}|u_m\rangle
\right]
=
2\text{Re}
\sum_{m}
M_{mn}^{(1)}(\vect{p}_c)
\vect{r}_{nm}(\vect{p}_c).
\label{W0W1}
\end{align}
Similarly, for $2\text{Re}\langle W_0|\vect{r}|W_2\rangle$, we find
\begin{align}
2\text{Re}\langle W_0|\vect{r}|W_2\rangle
=
2\text{Re}
\left[
\int_{\vect{p}} \int_{\vect{p}'}
\sum_{m \neq n}
M_{mn}^{(2)} (\vect{p})
C_n(\vect{p})C_n^*(\vect{p}')
e^{i(\vect{p}-\vect{p}') \cdot \vect{r}} \langle
u_n'|\vect{r}|u_m\rangle
\right]
=
2\text{Re}
\sum_{m}
M_{mn}^{(2)}(\vect{p}_c)
\vect{r}_{nm}(\vect{p}_c).
\label{W0W2}
\end{align}
Finally, for $\langle W_1|\vect{r}|W_1\rangle$, we find:
\begin{align}
&\langle W_1|\vect{r}|W_1\rangle
=
\int_{\vect{p}} \int_{\vect{p}'}
\sum_{m \neq n}
\sum_{l \neq n}
M_{mn}^{(1)} (\vect{p})
M_{ln}^{(1)*} (\vect{p}')
C_n(\vect{p})C_n^*(\vect{p}')
e^{i(\vect{p}-\vect{p}') \cdot \vect{r}} \langle
u_l'|\vect{r}|u_m\rangle
\nonumber \\
&=
\int_{\vect{p}} \int_{\vect{p}'}
\sum_{m \neq n}
\sum_{l \neq n}
M_{mn}^{(1)} (\vect{p})
M_{ln}^{(1)*} (\vect{p}')
C_n(\vect{p})C_n^*(\vect{p}')
[-i\delta_{ln}\partial_{\vect{p}}+\vect{\mathcal{A}}_{lm}]\delta(\vect{p}'-\vect{p})
\nonumber \\
&=
\int_{\vect{p}} \sum_{m \neq n}
M_{mn}^{(1)*} (\vect{p}) C_n^*(\vect{p})
i\partial_{\vect{p}} \left[ M_{mn}^{(1)}(\vect{p})C_n(\vect{p}) \right]
+
\sum_{l \neq n} \sum_{m \neq n} M_{ln}^{(1)*}(\vect{p}_c)\vect{\mathcal{A}}_{lm} (\vect{p}_c) M_{mn}^{(1)}(\vect{p}_c)
\nonumber \\
&=
\sum_{m \neq n} M_{mn}^{(1)*}(\vect{p}_c)i\partial_{\vect{p}_c} M_{mn}^{(1)}(\vect{p}_c)
+
\partial_{\vect{p}_c} \gamma(\vect{p}_c) \sum_{m \neq n} M_{mn}^{(1)*}(\vect{p}_c) M_{mn}^{(1)}(\vect{p}_c)
+
\sum_{l \neq n} \sum_{m \neq n} M_{ln}^{(1)*}(\vect{p}_c)\vect{\mathcal{A}}_{lm} (\vect{p}_c) M_{mn}^{(1)}(\vect{p}_c)
\nonumber \\
&+
\sum_{m \neq n} \int_{\vect{p}}
M_{mn}^{(1)*} (\vect{p}) M_{mn}^{(1)}(\vect{p})
\dfrac{i}{2}\partial_{\vect{p}}|C_n(\vect{p})|^2
\nonumber \\
&=
\sum_{m \neq n} M_{mn}^{(1)*}(\vect{p}_c)i\partial_{\vect{p}_c} M_{mn}^{(1)}(\vect{p}_c)
+
\partial_{\vect{p}_c} \gamma(\vect{p}_c) \sum_{m \neq n} M_{mn}^{(1)*}(\vect{p}_c) M_{mn}^{(1)}(\vect{p}_c)
+
\sum_{l \neq n} \sum_{m \neq n} M_{ln}^{(1)*}(\vect{p}_c)\vect{\mathcal{A}}_{lm} (\vect{p}_c) M_{mn}^{(1)}(\vect{p}_c)
\nonumber \\
&-
\sum_{m \neq n}
\dfrac{1}{2} i\partial_{\vect{p}_c}
\left[
M_{mn}^{(1)*} (\vect{p}_c) M_{mn}^{(1)}(\vect{p}_c)
\right]
\nonumber \\
&=
\text{Re} \sum_{m \neq n} M_{mn}^{(1)*}(\vect{p}_c)i\partial_{\vect{p}_c} M_{mn}^{(1)}(\vect{p}_c)
+
\partial_{\vect{p}_c} \gamma(\vect{p}_c) \sum_{m \neq n} M_{mn}^{(1)*}(\vect{p}_c) M_{mn}^{(1)}(\vect{p}_c)
+
\sum_{l \neq n} \sum_{m \neq n} M_{ln}^{(1)*}(\vect{p}_c)\vect{\mathcal{A}}_{lm} (\vect{p}_c) M_{mn}^{(1)}(\vect{p}_c).
\label{W1W1}
\end{align}
Collecting these results, we obtain (for simplicity, we will drop the explicit dependence on $\vect{p}_c$ below)
\begin{align}
\vect{r}_c
=
\partial_{\vect{p}_c}\gamma  + \vect{\mathcal{A}}_n  + \vect{a}_n^{(1)}
+
\vect{a}_n^{(2)} ,
\label{rc}
\end{align}
as given by Eq.(\ref{rcdef}) in the main text.
Here $\vect{a}_n^{(i)}$ is the $i$th \textit{positional shift} given by
\begin{align}
\vect{a}_n^{(1)}
\equiv
2\text{Re} \langle W_0|\vect{r}|W_1\rangle
=
2\text{Re} \sum_{m}
M_{mn}^{(1)}
\vect{r}_{nm}
\end{align}
and
\begin{align}
&\vect{a}_n^{(2)}
\equiv
2\delta\langle W_0|\vect{r}|W_0\rangle
+
2\text{Re} \langle W_0|\vect{r}|W_2\rangle
+
\langle W_1|\vect{r}|W_1\rangle
\nonumber \\
&=
- \sum_{m \neq n} M_{mn}^{(1)}  \vect{\mathcal{A}}_n M_{mn}^{(1)*}
+
2 \text{Re} \sum_{m \neq n}
M_{mn}^{(2)}
\vect{r}_{nm}
+
\text{Re} \sum_{m \neq n} M_{mn}^{(1)*}i\partial_{\vect{p}_c} M_{mn}^{(1)}
+
\sum_{l \neq n} \sum_{m \neq n} M_{ln}^{(1)*}\vect{\mathcal{A}}_{lm}  M_{mn}^{(1)}
\nonumber \\
&=
\text{Re} \sum_{m} M_{mn}^{(1)*}i\vect{\mathcal{D}}^{\vect{p}_c}_{mn} M_{mn}^{(1)}
+
\sum_{ml} M_{ln}^{(1)*}\vect{r}_{lm} M_{mn}^{(1)}
+
2 \text{Re} \sum_{m}
M_{mn}^{(2)}
\vect{r}_{nm}
\nonumber \\
&=
-
\text{Re} \sum_{m} M_{mn}^{(1)}i\vect{\mathcal{D}}^{\vect{p}_c}_{nm} M_{mn}^{(1)*}
+
\sum_{ml} M_{ln}^{(1)*}\vect{r}_{lm} M_{mn}^{(1)}
+
2 \text{Re} \sum_{m}
M_{mn}^{(2)}
\vect{r}_{nm}
\end{align}
where $2\delta \partial_{\vect{p}_c}\gamma$
is exactly canceled out with the second term of Eq.(\ref{W1W1}).
Furthermore, using the Levi-Civita symbol and Eq.(\ref{Mmn1}), we finally find:
\begin{align}
a_n^{(1;\lambda)}
&=
2\text{Re}
\sum_{m}
\sum_{\omega_1}
\dfrac{r^\lambda_{nm}r^\alpha_{mn}}{\omega_1-\epsilon_{mn}+i\xi} E_\alpha^{\omega_1},
\end{align}
where $E_\alpha^{\omega_1} \equiv e^{-i(\omega_1+i\xi)t}$,
as given by Eq.(\ref{an1}) in the main text.
Furthermore, by using Eq.(\ref{Mmn1}) and Eq.(\ref{Mmn2sol}), we find
\begin{align}
a_n^{(2;\lambda)}
&=
\text{Re}
\sum_{\omega_1 \omega_2}
\sum_{ml}
\dfrac{2r^\lambda_{nm}r^\beta_{ml}r^\alpha_{ln}}{(\omega_1-\epsilon_{ln}+i\xi)(\omega_{\Sigma}-\epsilon_{mn}+2i\xi)}
E_\alpha^{\omega_1} E_\beta^{\omega_2}
\nonumber \\
&+
\sum_{\omega_1 \omega_2}
\sum_{ml}
\dfrac{-r^\beta_{nl} r^\lambda_{lm} r^\alpha_{mn}}{(\omega_2-\epsilon_{nl}+i\xi)(\omega_1-\epsilon_{mn}+i\xi)}
E_\alpha^{\omega_1} E_\beta^{\omega_2}
\nonumber \\
&+
\text{Re}
\sum_{\omega_1 \omega_2}
\sum_m
\dfrac{2r^\lambda_{nm}}{\omega_{\Sigma}-\epsilon_{mn}+2i\xi}
i\mathcal{D}^\beta_{mn}
\left(
\dfrac{r^\alpha_{mn}}{\omega_1-\epsilon_{mn}+i\xi}
\right)
E_\alpha^{\omega_1} E_\beta^{\omega_2}
\nonumber \\
&+
\text{Re}
\sum_{\omega_1 \omega_2}
\sum_{m}
\dfrac{r^\beta_{mn}}{\omega_2-\epsilon_{mn}+i\xi}
i\mathcal{D}^\lambda_{nm}
\left(
\dfrac{r^\alpha_{nm}}{\omega_1-\epsilon_{nm}+i\xi}
\right)
E_\alpha^{\omega_1} E_\beta^{\omega_2}
\end{align}
as given by Eq.(\ref{an2}) in the main text, where we have used that
\begin{align}
M_{mn}^{(1)*}
=
\sum_{\omega_1} \dfrac{r^\alpha_{nm}E_\alpha e^{i(\omega_1-i\xi)t}}{\omega_1-\epsilon_{mn}-i\xi}
=
\sum_{\omega_1} \dfrac{r^\alpha_{nm}E_\alpha e^{-i(\omega_1+i\xi)t}}{-\omega_1-\epsilon_{mn}-i\xi}
=
-
M_{nm}^{(1)}
\end{align}
due to $\sum_{\omega_1=\pm\omega}f(\omega_1)=\sum_{\omega_1=\pm\omega}f(-\omega_1)$
for an arbitrary function $f(\omega_1)$.

\subsection{Eqs.(\ref{energy}-\ref{secondenergy})}
\label{wpenergy}

With the second-order wavepacket Eq.(\ref{wavepacket2}),
the energy for the wavepacket can be expressed as
\begin{align}
\bar{\epsilon}_n
&\equiv
\langle W_0|i\partial_t|W_0\rangle - \langle \bar{W} | (i\partial_t - H) |\bar{W}\rangle
\nonumber \\
&=
\left[
-
\langle W_0|i\partial_t\left(\delta|W_0\rangle \right) - \delta \langle W_0|i\partial_t|W_0\rangle
-
\langle W_1|i\partial_t|W_1\rangle
\right]
\nonumber \\
&+
\left[
\left(1+2\delta \right) \langle W_0|H_0|W_0\rangle
+
\langle W_1|H_0|W_1\rangle
+
\vect{E}\cdot\langle \bar{W}|\vect{r}|\bar{W}\rangle
\right]
+
\mathcal{O}(E^3)
\nonumber \\
&=
\left[
- i\partial_t \delta -2\delta \partial_t \gamma - \langle W_1|i\partial_t|W_1\rangle
\right]
+
\left[
(1+2\delta)\epsilon_n+\sum_{m} M_{mn}^{(1)}M_{mn}^{(1)*}\epsilon_{m} + \vect{E} \cdot \vect{r}_c
\right]
+
\mathcal{O}(E^3)
,
\label{energyapp}
\end{align}
where we have used Eq.(\ref{W0tW0}), $\langle W_0|W_1\rangle=0$, and $\langle W_0|W_2\rangle=0$.
Note that
\begin{align}
&\langle W_1|i\partial_t|W_1\rangle
=
\int_{\vect{p}} \int_{\vect{p}'}
\sum_{m \neq n} \sum_{l \neq n}
e^{i(\vect{p}-\vect{p}')\cdot\vect{r}} M_{ln}^{(1)*}(\vect{p}')
\langle u_l'|C_n(\vect{p}') i\partial_t \left(  M_{mn}^{(1)}(\vect{p}) C_n(\vect{p}) |u_m\rangle \right)
\nonumber \\
&=
\sum_{m} M_{mn}^{(1)} i\partial_t M_{mn}^{(1)*}
+
\int_{\vect{p}} \sum_{m} M_{mn}^{(1)*}(\vect{p})M_{mn}^{(1)}(\vect{p})
C_n(\vect{p}) i\partial_t C_n(\vect{p})
\nonumber \\
&=
\sum_{m} M_{mn}^{(1)} i\partial_t M_{mn}^{(1)*}
+
\partial_t \gamma \sum_{m} M_{mn}^{(1)}M_{mn}^{(1)*}
+
\int_{\vect{p}} \sum_{m} M_{mn}^{(1)*}(\vect{p})M_{mn}^{(1)}(\vect{p})
\dfrac{i}{2}\partial_t |C_n(\vect{p})|^2
\nonumber \\
&=
\sum_{m} M_{mn}^{(1)} i\partial_t M_{mn}^{(1)*}
+
\partial_t \gamma \sum_{m} M_{mn}^{(1)}M_{mn}^{(1)*}
+
\sum_{m}
\dfrac{i}{2} \partial_t
\left(
\int_{\vect{p}} M_{mn}^{(1)*}(\vect{p}) M_{mn}^{(1)}(\vect{p}) |C_n(\vect{p})|^2
\right)
\nonumber \\
&-
\sum_{m} \int_{\vect{p}}
\dfrac{i}{2} \partial_t
\left(
M_{mn}^{(1)*}(\vect{p}) M_{mn}^{(1)}(\vect{p})
\right)
|C_n(\vect{p})|^2
\nonumber \\
&=
\sum_{m} M_{mn}^{(1)} i\partial_t M_{mn}^{(1)*}
-
2\delta\partial_t\gamma,
\label{W1tW1app}
\end{align}
where we have used Eq.(\ref{delta}). Substituting Eq.(\ref{W1tW1app}) into Eq.(\ref{energyapp}), we find:
\begin{align}
\bar{\epsilon}_n
&=
\epsilon_n + \vect{E} \cdot \vect{r}_c
+
\dfrac{1}{2}i\partial_t \left(\sum_{m} M_{mn}^{(1)} M_{mn}^{(1)*} \right)
-
\sum_{m} M_{mn}^{(1)} i\partial_t M_{mn}^{(1)*}
+
\sum_{m} \epsilon_{mn} M_{mn}^{(1)} M_{mn}^{(1)*}
+
\mathcal{O}(E^3)
\nonumber \\
&=
\epsilon_n + \vect{E} \cdot \vect{r}_c
+
\dfrac{1}{2} \sum_{m}
\left[
\left(i\partial_t M_{mn}^{(1)}\right) M_{mn}^{(1)*}
-
M_{mn}^{(1)} i\partial_t M_{mn}^{(1)*}
\right]
+
\sum_{m} \epsilon_{mn} M_{mn}^{(1)} M_{mn}^{(1)*}
+
\mathcal{O}(E^3)
\nonumber \\
&=
\epsilon_n + \vect{E} \cdot \vect{r}_c
+
\text{Re}
\sum_{m}
\left[
\left(i\partial_t M_{mn}^{(1)}\right) M_{mn}^{(1)*}
+
\epsilon_{mn} M_{mn}^{(1)} M_{mn}^{(1)*}
\right]
\nonumber \\
&=
\epsilon_n + \vect{E}\cdot\vect{r}_c+\epsilon^{(2)}_n + \mathcal{O}(E^3),
\label{energyapp}
\end{align}
as given by Eq.(\ref{energy}) in the main text, where $\vect{r}_c \equiv \langle \bar{W}|\vect{r}|\bar{W}\rangle$
and
\begin{align}
\epsilon_n^{(2)}
&\equiv
\text{Re}
\sum_{m}
\left[
\left(i\partial_t M_{mn}^{(1)}\right) M_{mn}^{(1)*}
+
\epsilon_{mn} M_{mn}^{(1)} M_{mn}^{(1)*}
\right]
=
-\text{Re}\sum_{m} \sum_{\omega_1\omega_2}
\dfrac{(\omega_1+i\xi+\epsilon_{mn})r_{mn}^\alpha r^\beta_{nm} E_\alpha^{\omega_1}E_\beta^{\omega_2}}
{(\omega_1-\epsilon_{mn}+i\xi)(\omega_2-\epsilon_{nm}+i\xi)}
\nonumber \\
&=
-\text{Re}
\sum_{m} \sum_{\omega_1\omega_2}
\dfrac{r_{mn}^\alpha r^\beta_{nm} E_\alpha^{\omega_1}E_\beta^{\omega_2}}
{\omega_2-\epsilon_{nm}+i\xi}
=
-\dfrac{1}{2} \sum_{\omega_1} a_n^{(1;\alpha)} E_\alpha^{\omega_1},
\label{secondenergyapp}
\end{align}
as given by Eq.(\ref{secondenergy}) in the main text,
where we have used $M_{mn}^{(1)*}=-M_{nm}^{(1)}$ and Eq.(\ref{an1}).
Note that only the first term of Eq.(\ref{energyapp}) contains $\vect{r}_c$
while Eq.(\ref{secondenergyapp}) depends only on $\vect{k}_c$.

\section{The solution for Boltzmann equation}
\label{neqsol}

In this appendix, we will derive Eq.(\ref{neq1}) and Eq.(\ref{neq2}) given in the main text,
which are obtained by solving the Boltzmann equation in presence of the electric field.
For completeness, the Boltzmann equation is reproduced as follows:
\begin{equation}
\partial_t \bar{f}_n+\eta(\bar{f}_n-f_n^{(0)})=E_\alpha^{\omega_1}\partial_\alpha \bar{f}_n,
\label{boltzapp}
\end{equation}
where the relaxation time approximation $\tau=1/\eta$ is adopted.
Eq.(\ref{boltzapp}) can be iteratively solved from
the equilibrium distribution function $f_n^{(0)}=f_n=1/\left[ e^{(\epsilon_n-\mu)/k_BT}+1\right]$
without the electric field,
where $\mu$ is the chemical potential, $k_B$ is the Boltzmann constant,
and $T$ is the temperature. Straightfowardly,
by inserting $\bar{f}_n=f_n^{(0)}+f_n^{(1)}+f_n^{(2)}+\cdots$ into Eq.(\ref{boltzapp}),
where $f_n^{(i)} \propto E_\alpha^{(i)}$,
we obtain the following recursive first-order inhomogeneous linear differential equation equations
for the nonequilibrium distribution:
\begin{align}
\partial_t f_n^{(1)} + \eta f_n^{(1)} &= E_\alpha^{\omega_1} \partial_\alpha f_n^{(0)},
\label{OneOrder}
\\
\partial_t f_n^{(2)} + \eta f_n^{(2)} &= E_\beta^{\omega_2} \partial_\beta f_n^{(1)},
\label{TwoOrder}
\end{align}
which are truncated at the second order of the electric field.
Note that $f_n^{(i)}=0$ with $i=1,2$ in the absence of the electric field $(t \rightarrow -\infty)$
is the boundary condition for Eqs.(\ref{OneOrder}-\ref{TwoOrder}).
Recall that for the first-order inhomogeneous linear differential equation
$\partial_t y(t)+p(t)y(t)=q(t)$, the general solution is given by
\begin{equation}
y(t)=e^{-\int p(t)dt} \left( C+\int q(t) e^{\int p(t) dt}dt \right).
\label{general}
\end{equation}
Using Eq.(\ref{general}), for Eq.(\ref{OneOrder}), we immediately obtain
\begin{align}
f_n^{(1)}
=
e^{-\eta t}
\left(C+E_\alpha \partial_\alpha f_n^{(0)} \int e^{-i\left[\omega_1 +i(\eta+\xi)\right]t} dt \right)
=
\dfrac{E_\alpha^{\omega_1} \partial_\alpha f_n^{(0)} }{-i\left[\omega_1+i(\eta+\xi)\right]}
=
\dfrac{i\partial_\alpha f_n^{(0)} E_\alpha^{\omega_1}}{\omega_1+i(\eta+\xi)},
\label{neq1app}
\end{align}
where $p(t)=\eta$ and $q(t)=E_\alpha^{\omega_1}\partial_\alpha f_n^{(0)}$ in Eq.(\ref{general})
and $C=0$ fixed by the boundary condition.
Similarly, for Eq.(\ref{TwoOrder}), we obtain
\begin{align}
f_n^{(2)}
&=
e^{-\eta t}
\left(
C
+
\dfrac{i\partial_{\beta\alpha}^2f_n^{(0)}}{\omega_1+i(\eta+\xi)}
\int e^{-i\left[\omega_\Sigma+i(\eta+2\xi)\right]t} dt
\right)
=
\dfrac{E_\alpha^{\omega_1} E_\beta^{\omega_2} i\partial_{\alpha\beta}^2f_n^{(0)}}
{\left[\omega_1+i(\eta+\xi)\right]\left[-i\omega_\Sigma+(\eta+2\xi)\right]}
\nonumber \\
&=
\dfrac{-\partial_{\alpha\beta}^2f_n^{(0)} E_\alpha^{\omega_1} E_\beta^{\omega_2}}
{\left[\omega+i(\eta+\xi)\right]\left[\omega_\Sigma+i(\eta+2\xi)\right]},
\label{neq2app}
\end{align}
where $q(t)=f_n^{(1)}$ has been used in Eq.(\ref{general}). Eq.(\ref{neq1app})
and Eq.(\ref{neq2}) are given by Eq.(\ref{neq1}) and Eq.(\ref{neq2}) in the main text,
respectively.

\section{Response theory}
\label{responsesol}

In this appendix, we first derive the expressions for the density matrix element
by iteratively solving the quantum Liouville equation. With the density matrix element,
we further derive the linear and nonlinear conductivities given in the main text.

\subsection{The density matrix element}
The quantum Liouville equation for the density matrix in Bloch basis is reproduced as
\begin{align}
i\partial_t \rho_{mn}
=
[H,\rho]_{mn}
=
\epsilon_{mn}\rho_{mn}
+
[r^\alpha,\rho]_{mn}
E_\alpha^{\omega_1},
\label{rhomneq}
\end{align}
where $\epsilon_{mn}=\epsilon_m-\epsilon_n$ and
\begin{align}
[r^\alpha,\rho]_{mn}
=
i\mathcal{D}^{\alpha}_{mn}\rho_{mn}
+
\sum_{l}
\left(
r^\alpha_{ml}\rho_{ln}-\rho_{ml} r^\alpha_{ln}
\right).
\end{align}
Here $\mathcal{D}^\alpha_{mn} \equiv \partial_\alpha-i(\mathcal{A}_m^\alpha-\mathcal{A}_n^\alpha)$ is
the $U(1)$ covariant derivative.
Eq.(\ref{rhomneq}) can also be solved iteratively. Particularly, by writing
$\rho_{mn}=\sum_{i} \rho_{mn}^{(i)}$,
where $\rho_{mn}^{(i)} \propto E_\alpha^{(i)}$
and $\rho^{(0)}_{mn}=\delta_{mn}f_n$, we obtain the following
recursive first-order inhomogeneous linear differential equations
up to the second order of the electric field:
\begin{align}
\partial_t \rho_{mn}^{(1)} + i\epsilon_{mn}\rho_{mn}^{(1)}&=-i [r^\alpha,\rho^{(0)}]_{mn} E_\alpha^{\omega_1}
\label{rhomneq1}
\\
\partial_t \rho_{mn}^{(2)} + i\epsilon_{mn}\rho_{mn}^{(2)}&= -i[r^\beta,\rho^{(1)}]_{mn}E_\beta^{\omega_2},
\label{rhomneq2}
\end{align}
where the boundary condition is $\rho^{(i)}_{mn}(t)=0$ with $i \geq 1$
when $t \rightarrow -\infty $, which implies that the AC electric field is applied adiabatically.

For Eq.(\ref{rhomneq1}), by taking $p(t)=i\epsilon_{mn}$ and
\begin{align}
q(t)
=
-i[r^\alpha, \rho^{(0)}]_{mn}E_\alpha^{\omega_1}
=
\left[
\mathcal{D}^\alpha_{mn} \rho_{mn}^{(0)}
-
i
\sum_{l}
\left(r^\alpha_{ml} \rho_{ln}^{(0)}
-
\rho_{ml}^{(0)} r^\alpha_{ln}
\right)
\right]
E_\alpha^{\omega_1}
=
\left(\delta_{mn}\partial_\alpha f_m - i f_{nm}r^\alpha_{mn}\right)
E_\alpha^{\omega_1}
\end{align}
in Eq.(\ref{general}), where $f_{nm}=f_n-f_m$, we find
\begin{align}
\rho_{mn}^{(1)}
&=
e^{-i\epsilon_{mn}t}
\left[
C+
\left(
\delta_{mn}\partial_\alpha f_m - i f_{nm} r^\alpha_{mn}
\right)
E_\alpha
\int e^{-i ( \omega_1-\epsilon_{mn}+i\xi )t} dt
\right]
=
\dfrac{\delta_{mn}\partial_\alpha f_{m}-if_{nm}r^\alpha_{mn}}{-i(\omega_1-\epsilon_{mn}+i\xi)} E_\alpha^{\omega_1}
\nonumber \\
&=
\dfrac{i\delta_{mn}\partial_\alpha f_{m} E_\alpha^{\omega_1}}{\omega_1+i\xi}
+
\dfrac{f_{nm}r^\alpha_{mn} E_\alpha^{\omega_1}}{\omega_1-\epsilon_{mn}+i\xi}
,
\label{rhomneq1sol}
\end{align}
Note that $C=0$ is fixed by the boundary condition.
Similarly, for Eq.(\ref{rhomneq2}), we note that $q(t)=i\epsilon_{mn}$ and
\begin{align}
q(t)
&=
-i [r^\beta, \rho^{(1)}]_{mn}E_\beta^{\omega_2}
=
\left[
\mathcal{D}_{mn}^{\beta} \rho_{mn}^{(1)}
-
i\sum_{l}
\left(r^\beta_{ml} \rho_{ln}^{(1)}
-
\rho_{ml}^{(1)} r^\beta_{ln}
\right)
\right]
E_\beta^{\omega_2}
\nonumber \\
&=
\left[
\dfrac{ i\delta_{mn} \partial^2_{\alpha \beta}f_m }{\omega_1+i\xi}
+
\mathcal{D}^{\beta}_{mn}
\left(
\dfrac{f_{nm} r^\alpha_{mn} }{\omega_1-\omega_{mn}+i\xi}
\right)
\right]
E_\alpha^{\omega_1}
E_\beta^{\omega_2}
\nonumber \\
&
-
i
\left[
\sum_{l}
\left(
\dfrac{r^\beta_{ml}i\delta_{ln}\partial_\alpha f_l}{\omega_1+i\xi}
+
\dfrac{r^\beta_{ml}f_{nl}r^\alpha_{ln}}{\omega_1-\epsilon_{ln}+i\xi}
\right)
-
\sum_{l}
\left(
\dfrac{i\delta_{ml}\partial_\alpha f_m r^\beta_{ln}}{\omega_1+i\xi}
+
\dfrac{f_{lm}r^\alpha_{ml} r^\beta_{ln}}{\omega_1-\epsilon_{ml}+i\xi}
\right)
\right]
E_\alpha^{\omega_1}
E_\beta^{\omega_2}
\nonumber \\
&\equiv
T
E_\alpha^{\omega_1}
E_\beta^{\omega_2},
\end{align}
where
\begin{align}
T&=
\dfrac{ i\delta_{mn} \partial^2_{\alpha \beta}f_m}{\omega_1+i\xi}
+
\mathcal{D}^{\beta}_{mn}
\left(
\dfrac{f_{nm} r^\alpha_{mn}}{\omega_1-\omega_{mn}+i\xi}
\right)
+
\dfrac{r^\beta_{mn}\partial_\alpha f_{nm}}{\omega_1+i\xi}
-
i
\sum_{l}
\left(
\dfrac{r^\beta_{ml}r^\alpha_{ln}f_{nl}}{\omega_1-\epsilon_{ln}+i\xi}
-
\dfrac{f_{lm}r^\alpha_{ml}r^\beta_{ln}}{\omega_1-\epsilon_{ml}+i\xi}
\right).
\end{align}
Also by inserting the explicit expressions for $p(t)$ and $q(t)$ into Eq.(\ref{general}), we find:
\begin{align}
\rho^{(2)}_{mn}
&=
e^{-i\epsilon_{mn}t}
\left(C + T E_\alpha E_\beta \int e^{-i(\omega_\Sigma-\epsilon_{mn}+i2\xi)t} \right)
=
\dfrac{T E_\alpha^{\omega_1} E_\beta^{\omega_2}}{-i(\omega_\Sigma-\epsilon_{mn}+2i\xi)}
=
\dfrac{i T E_\alpha^{\omega_1} E_\beta^{\omega_2}}{\omega_\Sigma-\epsilon_{mn}+2i\xi}
\nonumber \\
&=
\left[
\dfrac{-\delta_{mn} \partial^2_{\alpha \beta} f_m}
{(\omega_1+i\xi)(\omega_\Sigma+2i\xi)}
+
\dfrac{1}{\omega_\Sigma-\epsilon_{mn}+2i\xi}
\mathcal{D}^{\beta}_{mn}
\left(
\dfrac{ i f_{nm} r^\alpha_{mn}}{\omega_1-\omega_{mn}+i\xi}
\right)
+
\dfrac{i r^\beta_{mn}\partial_\alpha f_{nm}}
{(\omega_1+i\xi)(\omega_\Sigma-\epsilon_{mn}+2i\xi)}
\right]
E_\alpha^{\omega_1}
E_\beta^{\omega_2}
\nonumber \\
&+
\sum_{l}
\dfrac{1}{\omega_\Sigma-\epsilon_{mn}+2i\xi}
\left(
\dfrac{r^\beta_{ml}r^\alpha_{ln}f_{nl}}{\omega_1-\epsilon_{ln}+i\xi}
-
\dfrac{f_{lm}r^\alpha_{ml}r^\beta_{ln}}{\omega_1-\epsilon_{ml}+i\xi}
\right)
E_\alpha^{\omega_1}E_\beta^{\omega_2}
,
\label{rhomneq2sol}
\end{align}
where $\omega_\Sigma=\omega_1+\omega_2$.
With Eq.(\ref{rhomneq1sol}) and Eq.(\ref{rhomneq2sol}), we are ready to calculate the
conductivity up to the second order.

\subsection{The conductivity}
In this subsection, we will derive Eqs.(\ref{sigmaDrude}-\ref{sigmaDisp})
and Eqs.(\ref{sigmaee}-\ref{sigmaii}) discussed in the main text.
In general, the current density can also be expressed as
$
J_\lambda (t) = \sum_{i=1}^{\infty} J^{(i)}_\lambda (t)
$,
where $J_\lambda^{(i)} (t) \propto E_\alpha^{(i)}$ particularly with
\begin{align}
J_\lambda^{(i)} \equiv \text{Tr}[\rho\hat{v}^\lambda] = \sum_{mn} \int_k \rho_{mn}^{(i)} v^{\lambda}_{nm}.
\label{currentith}
\end{align}
At the first order of the applied electric field, by substituting Eq.(\ref{rhomneq1sol}) into Eq.(\ref{currentith}),
we find that
$
J^{(1)}_\lambda= \sum_{mn} \int_k \rho_{mn}^{(1)} v^\lambda_{nm}
\equiv \sigma_{\lambda\alpha}^{(1)}(\omega_1) E_\alpha^{\omega_1}
$,
where the conductivity $\sigma_{\lambda\alpha}^{(1)}(\omega)$ is
\begin{align}
\sigma_{\lambda\alpha}^{(1)}(\omega_1)
&=
\sum_{n} \int_k
\dfrac{i v^\lambda_{n} \partial_\alpha f_n }{\omega_1+i\xi}
+
\sum_{mn} \int_k
\dfrac{f_{nm}r^\alpha_{mn}v^\lambda_{nm}}{\omega_1-\epsilon_{mn}+i\xi}
=
\sigma_{\lambda\alpha}^{(\text{1;1})}
+
\sum_{mn} \int_k
if_{nm}r^\alpha_{mn}r^\lambda_{nm}
\left(
1-
\dfrac{\omega_1+i\xi}{\omega_1-\epsilon_{mn}+i\xi} \right)
\nonumber \\
&=
\sigma_{\lambda\alpha}^{(\text{1;1})}
+
\sum_{n} \int_k f_n \Omega^{\lambda\alpha}_n
-
\sum_{nm} \int_k
\dfrac{i(\omega_1+i\xi)f_{nm}r^\alpha_{mn}r^\lambda_{nm}}{\omega_1-\epsilon_{mn}+i\xi}
=
\sigma_{\lambda\alpha}^{(1;1)}
+
\sigma_{\lambda\alpha}^{(1;0)}
+
\sigma_{\lambda\alpha}^{(1;\bar{1})},\label{eqC11}
\end{align}
as given by Eq.(\ref{sigma1}) in the main text.
Here we have used $v^\lambda_{nm}=i\epsilon_{nm}r^\lambda_{nm}$ for $m \neq n$
and $\epsilon_{nm}/(\omega_1-\epsilon_{mn}+i\xi)=1-(\omega_1+i\xi)/(\omega_1-\epsilon_{mn}+i\xi)$.
In addition, we have defined
$\Omega_{n}^{\lambda\alpha}=\sum_m i (r^\lambda_{nm}r^\alpha_{mn}-r^\alpha_{nm}r^\lambda_{mn})$ and
\begin{align}
\sigma_{\lambda\alpha}^{(1;1)}
=
\sum_{n} \int_k
\dfrac{i v^\lambda_{n} \partial_\alpha f_n }{\omega_1+i\xi},
\quad
\sigma_{\lambda\alpha}^{(1;0)}
=
\sum_{n} \int_k f_n \Omega^{\lambda\alpha}_n,
\quad
\sigma_{\lambda\alpha}^{(1;\bar{1})}
=
\sum_{nm} \int_k
\dfrac{-i(\omega_1+i\xi)f_{nm}r^\alpha_{mn}r^\lambda_{nm}}{\omega_1-\epsilon_{mn}+i\xi},
\end{align}
as given by Eqs.(\ref{sigmaDrude}-\ref{sigmaDisp}) in the main text.

Similarly, at the second order of the electric field,
by substituting Eq.(\ref{rhomneq2sol}) into Eq.(\ref{currentith}), we have
$
J^{(2)}_\lambda = \sum_{mn} \int_k \rho_{mn}^{(2)} v^\lambda_{nm}
\equiv
\sigma^{(2)}_{\lambda\alpha\beta}(\omega_1,\omega_2) E_\alpha^{\omega_1} E_\beta^{\omega_2}
$,
where
$
\sigma_{\lambda\alpha\beta}^{(2)}(\omega_1,\omega_2)
\equiv
\sigma_{\lambda\alpha\beta}^{(2,ee)}
+
\sigma_{\lambda\alpha\beta}^{(2,ei)}
+
\sigma_{\lambda\alpha\beta}^{(2,ie)}
+
\sigma_{\lambda\alpha\beta}^{(2,ii)}
$
with
\begin{align}
\sigma_{\lambda\alpha\beta}^{(2,ee)}
&=
\sum_{mnl} \int_k
\mathcal{V}^{\lambda}_{nm}
\left(\dfrac{r^\beta_{ml} r^\alpha_{ln}f_{nl} }{\omega_1-\epsilon_{ln}+i\xi}
-
\dfrac{f_{lm}r^\alpha_{ml}r^\beta_{ln}}{\omega_1 -\epsilon_{ml}+i\xi}
\right),
\\
\sigma_{\lambda\alpha\beta}^{(2,ei)}
&=
\sum_{mn} \int_k
\mathcal{V}^{\lambda}_{nm}
\left(\dfrac{ir^\beta_{mn}\partial_{\alpha}f_{nm}}{\omega_1+i\xi}\right),
\\
\sigma_{\lambda\alpha\beta}^{(2,ie)}
&=
\sum_{mn} \int_k
\mathcal{V}^{\lambda}_{nm}
\mathcal{D}_{mn}^\beta
\left(
\dfrac{if_{nm}r^\alpha_{mn}}{\omega_1-\epsilon_{mn}+i\xi}
\right),
\\
\sigma_{\lambda\alpha\beta}^{(2,ii)}
&=
\sum_{n} \int_k
\dfrac{-v^\lambda_{n}\partial^2_{\alpha \beta}f_n}{(\omega_1+i\xi)(\omega_\Sigma+2i\xi)},
\end{align}
as given by Eqs.(\ref{sigmaee}-\ref{sigmaii}) in the main text.
Here
$\mathcal{V}^{\lambda}_{nm} \equiv v^\lambda_{nm}/(\omega_\Sigma-\epsilon_{mn}+2i\xi)$.

\subsection{The inclusion of relaxation}
\label{modifyapp}

By phenomenologically including the relaxation into the quantum Liouville equation Eq. (\ref{QL0}), we find:
\begin{align}
\partial_t \rho_{mn}^{(1)} + (i\epsilon_{mn}+\eta)\rho_{mn}^{(1)}
&=
\left[
\mathcal{D}_{\alpha}^{mn}\rho_{mn}^{(0)}
-
i\sum_{l}
\left(
r^\alpha_{ml}\rho_{ln}^{(0)}-\rho_{ml}^{(0)} r^\alpha_{ln}
\right)
\right]
E_\alpha^{\omega_1},
\label{rhomneq1modify}
\\
\partial_t \rho_{mn}^{(2)} + (i\epsilon_{mn}+\eta)\rho_{mn}^{(2)}
&=
\left[
\mathcal{D}_{\beta}^{mn}\rho_{mn}^{(1)}
-
i\sum_{l}
\left(
r^\beta_{ml}\rho_{ln}^{(1)}-\rho_{ml}^{(1)} r^\beta_{ln}
\right)
\right]
E_\beta^{\omega_2},
\label{rhomneq2modify}
\end{align}
where $\eta=1/\tau$. As a result, by solving Eq.(\ref{rhomneq1modify}), we find
\begin{align}
\rho^{(1)}_{mn}(\omega_1)
=
\dfrac{\delta_{mn}\partial_\alpha f_{m}-if_{nm}r^\alpha_{mn}}{-i[\omega_1-\epsilon_{mn}+i(\xi+\eta)]} E_\alpha^{\omega_1}.
\end{align}
In the DC limit, this equation becomes
\begin{align}
\rho^{(1)}_{mn}
=
\tau \delta_{mn} \partial_\alpha f_m E_\alpha
+
\dfrac{f_{nm}r^\alpha_{mn}}{-\epsilon_{mn}+i\eta} E_\alpha,
\end{align}
where the second term gives the unphysical linear conductivity
\begin{align}
\sigma_{\lambda\alpha}^H=
\sum_{mn} \int_k
\dfrac{f_{nm}r^\alpha_{mn}v^\lambda_{nm}}{-\epsilon_{mn}+i\eta}
=
2 \sum_{mn} \int_k
f_n
\text{Re}
\dfrac{r^\alpha_{mn}v^\lambda_{nm}}{-\epsilon_{mn}+i/\tau},
\end{align}
as given by Eq.(\ref{L-Hall}) in the main text. However, if we further assume $\eta \rightarrow \bar{\eta}=\delta_{mn}/\tau$
and then we have
\begin{align}
\rho^{(1)}_{mn}(\omega_1)
=
\dfrac{i\delta_{mn}\partial_\alpha f_{m} E_\alpha^{\omega_1}}{\omega_1+i(\eta+\xi)}
+
\dfrac{f_{nm}r^\alpha_{mn} E_\alpha^{\omega_1}}{\omega_1-\epsilon_{mn}+i\xi}
\label{rhomn1solmodify}
\end{align}
which gives the same linear current as the semiclassical theory in both AC and DC regimes. Furthermore,
by substituting Eq.(\ref{rhomn1solmodify}) into Eq.(\ref{rhomneq2modify}),
where $\eta$ is also replaced by $\bar{\eta}$, we find
\begin{align}
\rho^{(2)}_{mn}
&=
\dfrac{-\delta_{mn} \partial^2_{\alpha \beta} f_m}
{\left[\omega_1+i(\eta+\xi)\right]\left[\omega_\Sigma+i(\eta+2\xi)\right]}
E_\alpha^{\omega_1} E_\beta^{\omega_2}
+
\dfrac{1}{\omega_\Sigma-\epsilon_{mn}+i2\xi}
\mathcal{D}_{\beta}^{mn}
\left(
\dfrac{ i f_{nm} r^\alpha_{mn}}{\omega_1-\omega_{mn}+i\xi}
\right)
E_\alpha^{\omega_1} E_\beta^{\omega_2}
\nonumber \\
&+
\dfrac{1}{\omega_\Sigma-\epsilon_{mn}+i2\xi}
\dfrac{i r^\beta_{mn}\partial_\alpha f_{nm}}
{\omega_1+i(\eta+\xi)}
E_\alpha^{\omega_1}E_\beta^{\omega_2}
+
\sum_{l}
\left(
\dfrac{1}{\omega_\Sigma-\epsilon_{mn}+i(\bar{\eta}+2\xi)}
\right)
\left(
\dfrac{r^\beta_{ml}r^\alpha_{ln}f_{nl}}{\omega_1-\epsilon_{ln}+i\xi}
-
\dfrac{f_{lm}r^\alpha_{ml}r^\beta_{ln}}{\omega_1-\epsilon_{ml}+i\xi}
\right)
\label{rhomneq2solmodify}
\end{align}
where $\omega_\Sigma=\omega_1+\omega_2$.
Straightforwardly, substituting Eq.(\ref{rhomneq2solmodify}) into the definition of current density
$J^{(2)}_\lambda = \sum_{mn} \int_k \rho_{mn}^{(2)} v^\lambda_{nm}
\equiv
\sigma^{(2)}_{\lambda\alpha\beta}(\omega_1,\omega_2) E_\alpha^{\omega_1} E_\beta^{\omega_2}$,
we find
\begin{align}
\sigma_{\lambda\alpha\beta}^{(2)}(\omega_1,\omega_2)
&=
\sigma_{\lambda\alpha\beta}^{(2,ee)}
+
\sigma_{\lambda\alpha\beta}^{(2,ei)}
+
\sigma_{\lambda\alpha\beta}^{(2,ie)}
+
\sigma_{\lambda\alpha\beta}^{(2,ii)}
\end{align}
with
\begin{align}
\sigma_{\lambda\alpha\beta}^{(2,ee)}
&=
\sum_{mnl}
\int_k
\dfrac{v^\lambda_{nm} }{\omega_\Sigma-\epsilon_{mn}+i(\bar{\eta}+2\xi)}
\left(\dfrac{r^\beta_{ml} r^\alpha_{ln}f_{nl} }{\omega_1-\epsilon_{ln}+i\xi}
-
\dfrac{f_{lm}r^\alpha_{ml}r^\beta_{ln}}{\omega_1 -\epsilon_{ml}+i\xi}
\right),
\\
\sigma_{\lambda\alpha\beta}^{(2,ei)}
&=
\sum_{mn}
\int_k
\left(\dfrac{v^\lambda_{nm}}{\omega_\Sigma-\epsilon_{mn}+i2\xi}\right)
\left(\dfrac{ir^\beta_{mn}\partial_{\alpha}f_{nm}}{\omega_1+i(\eta+\xi)}\right),
\\
\sigma_{\lambda\alpha\beta}^{(2,ie)}
&=
\sum_{mn} \int_k
\left(\dfrac{v^\lambda_{nm}}{\omega_\Sigma-\epsilon_{mn}+2i\xi}\right)
\mathcal{D}^{mn}_\beta
\left(
\dfrac{if_{nm}r^\alpha_{mn}}{\omega_1-\epsilon_{mn}+i\xi}
\right),
\\
\sigma_{\lambda\alpha\beta}^{(2,ii)}
&=
\sum_{n} \int_k
\dfrac{-v^\lambda_{n}\partial^2_{\alpha \beta}f_n}{\left[\omega_1+i(\eta+\xi)\right]\left[\omega_\Sigma+i(\eta+2\xi)\right]},
\end{align}
as given by Eqs. (\ref{sigmaeemodify2}-\ref{sigmaiimodify2}) in the main text.

\section{Calculation details in Section \ref{equivalence}}

\subsection{Eq.(\ref{sigmaee1})}
\label{sigmaeemeqn}

From Eq.(\ref{sigmaee}), when $m=n$ we find
\begin{align}
\sigma^{(2;ee)}_{\lambda\alpha\beta}
(m=n)
&\equiv
\sum_{nl}
\int_k
\dfrac{v^\lambda_{n}}{(\omega_\Sigma+i2\xi)}
\left(
\dfrac{r^\beta_{nl}r^\alpha_{ln}f_{nl}}{\omega_1-\epsilon_{ln}+i\xi}
-
\dfrac{f_{ln}r^\alpha_{nl}r^\beta_{ln}}{\omega_1-\epsilon_{nl}+i\xi}
\right)
\nonumber \\
&=
\sum_{nm}
\int_k
\dfrac{v^\lambda_{n} f_{nm}}{(\omega_\Sigma+i2\xi)}
\left(
\dfrac{r^\beta_{nm}r^\alpha_{mn}}{\omega_1-\epsilon_{mn}+i\xi}
+
\dfrac{r^\alpha_{nm}r^\beta_{mn}}{\omega_1-\epsilon_{nm}+i\xi}
\right)
\nonumber \\
&=
\dfrac{1}{2}
\sum_{nm}
\int_k
\dfrac{v^\lambda_{n} f_{nm}}{(\omega_\Sigma+i2\xi)}
\left(
\dfrac{r^\beta_{nm}r^\alpha_{mn}}{\omega_1-\epsilon_{mn}+i\xi}
+
\dfrac{r^\alpha_{nm}r^\beta_{mn}}{\omega_1-\epsilon_{nm}+i\xi}
+
\dfrac{r^\alpha_{nm}r^\beta_{mn}}{\omega_2-\epsilon_{mn}+i\xi}
+
\dfrac{r^\beta_{nm}r^\alpha_{mn}}{\omega_2-\epsilon_{nm}+i\xi}
\right)
\nonumber \\
&=
\dfrac{1}{2}
\sum_{nm}
\int_k
\left[
\dfrac{v^\lambda_{n} f_{nm}r^\beta_{nm}r^\alpha_{mn}}
{\left( \omega_1-\epsilon_{mn} + i\xi \right)\left( \omega_2+\epsilon_{mn} + i\xi \right)}
+
\dfrac{v^\lambda_{n} f_{nm}r^\alpha_{nm}r^\beta_{mn}}
{\left( \omega_1+\epsilon_{mn} + i\xi \right)\left( \omega_2-\epsilon_{mn} + i \xi \right)}
\right]
\nonumber \\
&=
\dfrac{1}{2}
\sum_{nm}
\int_k
\dfrac{\Delta^\lambda_{nm} f_{nm}r^\beta_{nm}r^\alpha_{mn}}
{\left( \omega_1-\epsilon_{mn} + i\xi \right)\left( \omega_2+\epsilon_{mn} + i\xi \right)}
\equiv
\sigma_{\lambda\alpha\beta}^{(2;ee1)},\label{eqD1}
\end{align}
where the third line is obtained by symmetrizing $(\alpha,\omega_1)$ and $(\beta,\omega_2)$
and $\Delta^\lambda_{nm}=v_n^\lambda-v_m^\lambda$ in the final result. In addition, from the third line
of Eq.(\ref{eqD1}) to the fourth line, we have used the following relation to remove the divergence,
\begin{align}
\dfrac{1}{\omega_1-\epsilon_{nm}+i\xi}
+
\dfrac{1}{\omega_2-\epsilon_{mn}+i\xi}
&=
\dfrac{\omega_\Sigma+ 2i\xi}
{\left( \omega_1-\epsilon_{mn} + i\xi \right)\left( \omega_2+\epsilon_{mn} + i\xi \right)}\label{eqD2}
\end{align}

So Eq.(\ref{eqD1}) is the same as Eq.(\ref{sigmaee1}) in the main text.

\subsection{Eq.(\ref{J20eeeiie})}
\label{intrinsicAC}

To prove Eq.(\ref{J20eeeiie}), we first define $J_\lambda^{(2;0)}=D+A$, where
\begin{align}
D
&\equiv
\sum_n \int_k f_n \left( v_n^{(2;\lambda)} + \Omega_n^{(1;\lambda\alpha)}E_\alpha^{\omega_1} \right)
=
\sum_n \int_k f_n
\left(
\dfrac{1}{2} \partial_\lambda a_n^{(1;\alpha)} - \partial_\alpha a_n^{(1;\lambda)}
\right)
E_\alpha^{\omega_1}
\nonumber \\
&=
-
\sum_{mn} \int_k
\left(
\dfrac{\partial_\lambda f_{nm} r^\beta_{nm}r^\alpha_{mn}}{2(\omega_1-\epsilon_{mn}+i\xi)}
-
\dfrac{\partial_\alpha f_{nm} r^\lambda_{nm}r^\beta_{mn}}{\omega_2-\epsilon_{mn}+i\xi}
\right)
E_\alpha^{\omega_1}
E_\beta^{\omega_2}
\nonumber \\
&=
-
\sum_{mn} \int_k
\left(
\dfrac{\partial_\lambda f_{nm} r^\beta_{nm}r^\alpha_{mn}}{2(\omega_1-\epsilon_{mn}+i\xi)}
-
\dfrac{\partial_\beta f_{nm} r^\lambda_{nm}r^\alpha_{mn}}{\omega_1-\epsilon_{mn}+i\xi}
\right)
E_\alpha^{\omega_1}
E_\beta^{\omega_2}
\label{Dresult}
,
\end{align}
which restores Eq.(\ref{J20DC}) in the DC limit after symmetrizing $\alpha$ and $\beta$,
as expected. In addition, we have
\begin{align}
A \equiv \sum_n \int_k f_n \partial_t a_n^{(2;\lambda)} = A_1+A_2+A_3
\end{align}
with
\begin{align}
A_1
&=
\text{Re}
\sum_{\omega_1\omega_2}
\sum_{nml}
\int_k f_n
\dfrac{-2i(\omega_\Sigma+2i\xi)r^\lambda_{nm}r^\beta_{ml}r^\alpha_{ln} E_\alpha^{\omega_1} E_\beta^{\omega_2}}
{(\omega_1-\epsilon_{ln}+i\xi)(\omega_{\Sigma}-\epsilon_{mn}+2i\xi)}
+
\sum_{\omega_1\omega_2}
\sum_{nml}
\int_k f_n
\dfrac{i(\omega_\Sigma+2i\xi)r^\beta_{nl} r^\lambda_{lm} r^\alpha_{mn}E_\alpha^{\omega_1} E_\beta^{\omega_2}}
{(\omega_2-\epsilon_{nl}+i\xi)(\omega_1-\epsilon_{mn}+i\xi)}
\\
A_2
&=
\text{Re}
\sum_{\omega_1\omega_2}
\sum_{nm}
\int_k f_n
\dfrac{2(\omega_\Sigma+2i\xi)r^\lambda_{nm}}{\omega_{\Sigma}-\epsilon_{mn}+2i\xi}
\mathcal{D}^\beta_{mn}
\left(
\dfrac{r^\alpha_{mn}}{\omega_1-\epsilon_{mn}+i\xi}
\right)
E_\alpha^{\omega_1}E_\beta^{\omega_2}
\\
A_3
&=
\text{Re}
\sum_{\omega_1\omega_2}
\sum_{nm}
\int_k f_n
\dfrac{-(\omega_\Sigma+2i\xi)r^\beta_{nm} }{\omega_2-\epsilon_{nm}+i\xi}
\mathcal{D}^\lambda_{mn}
\left(
\dfrac{r^\alpha_{mn}}{\omega_1-\epsilon_{mn}+i\xi}
\right)
E_\alpha^{\omega_1} E_\beta^{\omega_2},
\end{align}
where we have used Eq.(\ref{an2}) and
$\partial_t (E_\alpha^{\omega_1}E_\beta^{\omega_2})=-i(\omega_\Sigma+i2\xi)E_\alpha^{\omega_1}E_\beta^{\omega_2}$.
Note that $A_i$ with $i=1,2,3$ vanishes in the DC limit.

For $A_1$, using
\begin{align}
\dfrac{\omega_\Sigma+2i\xi}{\omega_\Sigma-\epsilon_{mn}+2i\xi}
=
1-\dfrac{\epsilon_{nm}}{\omega_\Sigma-\epsilon_{mn}+2i\xi},
\label{sumfact}
\end{align}
and
\begin{align}
\dfrac{\omega_\Sigma+i2\xi}{(\omega_2-\epsilon_{nl}+i\xi)(\omega_1-\epsilon_{mn}+i\xi)}
=
\dfrac{1}{\omega_2-\epsilon_{nl}+i\xi}
+
\dfrac{1}{\omega_1-\epsilon_{mn}+i\xi}
-
\dfrac{\epsilon_{lm}}{(\omega_2-\epsilon_{nl}+i\xi)(\omega_1-\epsilon_{mn}+i\xi)}
\end{align}
we find
\begin{align}
A_1
&=
2\text{Re}
\sum_{nml}
\int_k f_n
\left[
\dfrac{ - i r^\lambda_{nm} r_{ml}^\beta r^\alpha_{ln} }
{\omega_1-\epsilon_{ln}+i\xi}
+
\dfrac{ i\epsilon_{nm} r^\lambda_{nm} r_{ml}^\beta r^\alpha_{ln} }
{(\omega_1-\epsilon_{ln}+i\xi)(\omega_\Sigma-\epsilon_{mn}+2i\xi)}
\right]
E_\alpha^{\omega_1}
E_\beta^{\omega_2}
\nonumber \\
&+
\sum_{nml} \int_k
f_n
\left[
\dfrac{ir^\beta_{nl}r^\lambda_{lm}r^\alpha_{mn}}{\omega_2-\epsilon_{nl}+i\xi}
+
\dfrac{ir^\beta_{nl}r^\lambda_{lm}r^\alpha_{mn}}{\omega_1-\epsilon_{mn}+i\xi}
-
\dfrac{i\epsilon_{lm}r^\beta_{nl}r^\lambda_{lm}r^\alpha_{mn}}
{(\omega_2-\epsilon_{nl}+i\xi)(\omega_1-\epsilon_{mn}+i\xi)}
\right]
E_\alpha^{\omega_1}
E_\beta^{\omega_2}.
\label{A1temp1}
\end{align}
Here the first term of the first line and the first two terms of the second line
(denoted as $A_1^1$) can be simplified as
\begin{align}
A_1^1
&=
\sum_{\omega_1\omega_2}
\sum_{nml}
\int_k f_n
\left[
\dfrac{ - i r^\lambda_{nm} r_{ml}^\beta r^\alpha_{ln} }
{\omega_1-\epsilon_{ln}+i\xi}
+
\dfrac{ + i r^\lambda_{mn} r_{lm}^\beta r^\alpha_{nl} }
{\omega_1-\epsilon_{ln}-i\xi}
+
\dfrac{ir^\beta_{nl}r^\lambda_{lm}r^\alpha_{mn}}{\omega_2-\epsilon_{nl}+i\xi}
+
\dfrac{ir^\beta_{nl}r^\lambda_{lm}r^\alpha_{mn}}{\omega_1-\epsilon_{mn}+i\xi}
\right]
E_\alpha^{\omega_1}
E_\beta^{\omega_2}
\nonumber \\
&=
\sum_{\omega_1\omega_2}
\sum_{nml}
\int_k
\dfrac{ir^\alpha_{mn}}{\omega_1-\epsilon_{mn}+i\xi}
\left[
-f_n r^\lambda_{nl} r_{lm}^\beta
-
f_m r^\beta_{nl}r^{\lambda}_{lm}
+
f_m r^\lambda_{nl}r^{\beta}_{lm}
+
f_n r^\beta_{nl}r^{\lambda}_{lm}
\right]
E_\alpha^{\omega_1}
E_\beta^{\omega_2}
\nonumber \\
&=
\sum_{\omega_1\omega_2}
\sum_{nm}
\int_k
\dfrac{ir^\alpha_{mn}f_{nm}}{\omega_1-\epsilon_{mn}+i\xi}
\sum_l
\left(
r^\beta_{nl}r^{\lambda}_{lm}
-
r^\lambda_{nl}r^{\beta}_{lm}
\right)
E_\alpha^{\omega_1}
E_\beta^{\omega_2}
\nonumber \\
&=
\sum_{\omega_1\omega_2}
\sum_{nm}
\int_k
\dfrac{ir^\alpha_{mn}f_{nm}}{\omega_1-\epsilon_{mn}+i\xi}
\left(
\mathcal{D}^\beta_{nm} r^\lambda_{nm}
-
\mathcal{D}^\lambda_{nm} r^\beta_{nm}
\right)
E_\alpha^{\omega_1}
E_\beta^{\omega_2}
.
\end{align}
Note that to obtain the second and third terms before interchanging the dummy indices
we have used the $\sum_{\omega_1=\pm\omega}f(\omega_1)=\sum_{\omega_1=\pm\omega}f(-\omega_1)$
and interchanged $(\alpha,\omega_1)$ and $(\beta,\omega_2)$, respectively.
And Eq.(\ref{identity}) is used to obtain the final line.
For the remaining two terms of in Eq.(\ref{A1temp1}) (denoted as $A_1^2$), we find
\begin{align}
A_1^2
&=
\sum_{\omega_1\omega_2}
\sum_{nml}^{n \neq m}
\int_k f_n
\left[
\dfrac{ v^\lambda_{nm} r_{ml}^\beta r^\alpha_{ln} }
{(\omega_1-\epsilon_{ln}+i\xi)(\omega_\Sigma-\epsilon_{mn}+2i\xi)}
+
\dfrac{ v^\lambda_{mn} r_{lm}^\beta r^\alpha_{nl} }
{(\omega_1-\epsilon_{ln}-i\xi)(\omega_\Sigma-\epsilon_{mn}-2i\xi)}
\right]
E_\alpha^{\omega_1}
E_\beta^{\omega_2}
\nonumber \\
&-
\sum_{\omega_1\omega_2}
\sum_{nml}^{l \neq m} \int_k
f_n
\dfrac{r^\beta_{nl}v^\lambda_{lm}r^\alpha_{mn}}
{(\omega_2-\epsilon_{nl}+i\xi)(\omega_1-\epsilon_{mn}+i\xi)}
E_\alpha^{\omega_1}
E_\beta^{\omega_2},
\end{align}
where we have used $v^\lambda_{nm}=i\epsilon_{nm}r^\lambda_{nm}$.
By interchanging $l$ and $n$ of the last term and using
\begin{align}
\dfrac{1}{(\omega_2-\epsilon_{ln}+i\xi)(\omega_1-\epsilon_{ml}+i\xi)}
&=
\dfrac{1}{\omega_\Sigma-\epsilon_{mn}+2i\xi}
\left(
\dfrac{1}{\omega_1-\epsilon_{ml}+i\xi}
+
\dfrac{1}{\omega_2-\epsilon_{ln}+i\xi}
\right),
\end{align}
we further obtain
\begin{align}
A_1^2
&=
\sum_{\omega_1\omega_2}
\sum_{nml}^{n \neq m}
\int_k
\dfrac{v^\lambda_{nm}}{\omega_\Sigma-\epsilon_{mn}+2i\xi}
\left(
\dfrac{f_n r_{ml}^\beta r^\alpha_{ln} }
{\omega_1-\epsilon_{ln}+i\xi}
+
\dfrac{ f_m r_{ln}^\beta r^\alpha_{ml} }
{\omega_1+\epsilon_{lm}+i\xi}
\right)
E_\alpha^{\omega_1}
E_\beta^{\omega_2}
\nonumber \\
&-
\sum_{\omega_1\omega_2}
\sum_{nml}^{n \neq m} \int_k
\dfrac{v^\lambda_{nm}}{\omega_\Sigma-\epsilon_{mn}+2i\xi}
\left(
\dfrac{f_l r^\beta_{ln}r^\alpha_{ml}}
{\omega_1-\epsilon_{ml}+i\xi}
+
\dfrac{f_l r^\beta_{ln}r^\alpha_{ml}}
{\omega_2-\epsilon_{ln}+i\xi}
\right)
E_\alpha^{\omega_1}
E_\beta^{\omega_2}
\nonumber \\
&=
\sum_{\omega_1\omega_2}
\sum_{nml}^{n \neq m}
\int_k
\dfrac{v^\lambda_{nm}}{\omega_\Sigma-\epsilon_{mn}+2i\xi}
\left(
\dfrac{f_{nl} r_{ml}^\beta r^\alpha_{ln} }
{\omega_1-\epsilon_{ln}+i\xi}
+
\dfrac{ f_{ml}r^\alpha_{ml} r_{ln}^\beta }
{\omega_1-\epsilon_{ml}+i\xi}
\right)
E_\alpha^{\omega_1}
E_\beta^{\omega_2}
\nonumber \\
&=
\sum_{\omega_1\omega_2}
\sigma^{(2;ee)}_{\lambda\alpha\beta}(m \neq n)
E_\alpha^{\omega_1}
E_\beta^{\omega_2},
\end{align}
where $\sigma^{(2;ee)}_{\lambda\alpha\beta}$ is defined in Eq.(\ref{sigmaee}).
As a result, we find
\begin{align}
A_1=A_1^2+A_1^2 =
\sum_{\omega_1\omega_2}
\sum_{nm}
\int_k
\dfrac{ir^\alpha_{mn}f_{nm}}{\omega_1-\epsilon_{mn}+i\xi}
\left(
\mathcal{D}^\beta_{nm} r^\lambda_{nm}
-
\mathcal{D}^\lambda_{nm} r^\beta_{nm}
\right)
E_\alpha^{\omega_1}
E_\beta^{\omega_2}
+
\sum_{\omega_1\omega_2}
\sigma^{(2;ee)}_{\lambda\alpha\beta}(m \neq n)
E_\alpha^{\omega_1}
E_\beta^{\omega_2}.
\label{A1result}
\end{align}

For $A_2$, we find
\begin{align}
A_2
&=
\sum_{\omega_1\omega_2}
\sum_{nm}
\int_k f_n
\left[
\dfrac{(\omega_\Sigma+2i\xi)r^\lambda_{nm}}{\omega_{\Sigma}-\epsilon_{mn}+2i\xi}
\mathcal{D}^\beta_{mn}
\left(
\dfrac{r^\alpha_{mn}}{\omega_1-\epsilon_{mn}+i\xi}
\right)
+
\dfrac{(\omega_\Sigma-2i\xi)r^\lambda_{mn}}{\omega_{\Sigma}-\epsilon_{mn}-2i\xi}
\mathcal{D}^\beta_{nm}
\left(
\dfrac{r^\alpha_{nm}}{\omega_1-\epsilon_{mn}-i\xi}
\right)
\right]
E_\alpha^{\omega_1}E_\beta^{\omega_2}
\nonumber \\
&=
\sum_{\omega_1\omega_2}
\sum_{nm}
\int_k f_{nm}
\dfrac{(\omega_\Sigma+2i\xi)r^\lambda_{nm}}{\omega_{\Sigma}-\epsilon_{mn}+2i\xi}
\mathcal{D}^\beta_{mn}
\left(
\dfrac{r^\alpha_{mn}}{\omega_1-\epsilon_{mn}+i\xi}
\right)
\end{align}
where we have used the relation
$
\sum_{\omega_1=\pm\omega}\sum_{\omega_2=\pm\omega}f(\omega_1,\omega_2)
=
\sum_{\omega_1=\pm\omega}\sum_{\omega_2=\pm\omega}f(-\omega_1,-\omega_2)
$
for the second term. Furthermore, using Eq.(\ref{sumfact}), we find
\begin{align}
A_2
&=
\sum_{\omega_1\omega_2}
\sum_{nm}
\int_k
\left(
-
\dfrac{r^\lambda_{nm}r^\alpha_{mn}\partial_\beta f_{nm}}{\omega_1-\epsilon_{mn}+i\xi}
-
\dfrac{r^\alpha_{mn} f_{nm}}{\omega_1-\epsilon_{mn}+i\xi}
\mathcal{D}^\beta_{nm} r^\lambda_{nm}
\right)
E_\alpha^{\omega_1}E_\beta^{\omega_2}
\nonumber \\
&+
\sum_{\omega_1\omega_2}
\sum_{nm}
\int_k
\left[
\dfrac{v^\lambda_{nm}}{\omega_\Sigma-\epsilon_{mn}+2i\xi}
\mathcal{D}_{mn}^\beta
\left(
\dfrac{if_{nm}r^\alpha_{mn}}{\omega_1-\epsilon_{mn}+i\xi}
\right)
-
\dfrac{iv^\lambda_{nm} r^\alpha_{mn} \partial_\beta f_{nm}}{(\omega_\Sigma-\epsilon_{mn}+2i\xi)(\omega_1-\epsilon_{mn}+i\xi)}
\right]
E_\alpha^{\omega_1}E_\beta^{\omega_2}
\nonumber \\
&=
\sum_{\omega_1\omega_2}
\sum_{nm}
\int_k
\left(
-
\dfrac{r^\lambda_{nm}r^\alpha_{mn}\partial_\beta f_{nm}}{\omega_1-\epsilon_{mn}+i\xi}
-
\dfrac{r^\alpha_{mn} f_{nm}}{\omega_1-\epsilon_{mn}+i\xi}
\mathcal{D}^\beta_{nm} r^\lambda_{nm}
\right)
E_\alpha^{\omega_1}E_\beta^{\omega_2}
+
\left(
\sigma^{(2;ie)}_{\lambda\alpha\beta}
+
\sigma^{(ei;0)}_{\lambda\alpha\beta}
\right)
E_\alpha^{\omega_1}E_\beta^{\omega_2},
\label{A2result}
\end{align}
where we have used Eq.(\ref{sigmaie}) and Eq.(\ref{sigmaei0}).

For $A_3$, using integration by parts we find:
\begin{align}
A_3
&=
\text{Re}
\sum_{\omega_1\omega_2}
\sum_{nm}
\int_k
\left[
\dfrac{(\omega_\Sigma+2i\xi) f_n r^\alpha_{mn} \mathcal{D}^\lambda_{nm} r^\beta_{nm} }
{(\omega_2-\epsilon_{nm}+i\xi)(\omega_1-\epsilon_{mn}+i\xi)}
+
\dfrac{r^\beta_{nm} r^\alpha_{mn}}{\omega_1-\epsilon_{mn}+i\xi}
\partial_\lambda
\left(
\dfrac{(\omega_\Sigma+2i\xi) f_n }{\omega_2-\epsilon_{nm}+i\xi}
\right)
\right]
E_\alpha^{\omega_1} E_\beta^{\omega_2}
\\
&=
\text{Re}
\sum_{\omega_1\omega_2}
\sum_{nm}
\int_k
\left[
\dfrac{r^\alpha_{mn} \mathcal{D}^\lambda_{nm} r^\beta_{nm} }
{\omega_1-\epsilon_{mn}+i\xi}
+
\dfrac{r^\alpha_{mn} \mathcal{D}^\lambda_{nm} r^\beta_{nm} }
{\omega_2-\epsilon_{nm}+i\xi}
\right]
f_n
E_\alpha^{\omega_1} E_\beta^{\omega_2}
\nonumber \\
&+
\text{Re}
\sum_{\omega_1\omega_2}
\sum_{nm}
\int_k
\left[
\dfrac{r^\alpha_{mn} r^\beta_{nm} }
{\omega_1-\epsilon_{mn}+i\xi}
+
\dfrac{r^\alpha_{mn} r^\beta_{nm} }
{\omega_2-\epsilon_{nm}+i\xi}
\right]
\partial_\lambda f_n
E_\alpha^{\omega_1} E_\beta^{\omega_2}
\nonumber \\
&+
\text{Re}
\sum_{\omega_1\omega_2}
\sum_{nm}
\int_k
\left[
\dfrac{r^\beta_{nm}r^\alpha_{mn}\Delta^\lambda_{nm}}{(\omega_1-\epsilon_{mn}+i\xi)(\omega_2-\epsilon_{nm}+i\xi)}
+
\dfrac{r^\beta_{nm}r^\alpha_{mn}\Delta^\lambda_{nm}}{(\omega_2-\epsilon_{nm}+i\xi)^2}
\right]
f_n
E_\alpha^{\omega_1} E_\beta^{\omega_2}
\nonumber \\
&=
\sum_{\omega_1\omega_2} \sum_{mn} \int_k
\dfrac{f_{nm}r^\alpha_{mn}\mathcal{D}^\lambda_{nm}r^\beta_{nm}}{\omega_1-\epsilon_{mn}+i\xi}
E_\alpha^{\omega_1} E_\beta^{\omega_2}
+
\dfrac{1}{2}
\sum_{\omega_1\omega_2} \sum_{mn} \int_k
\dfrac{\partial_\lambda f_{nm} r^\alpha_{mn}r^\beta_{nm}}{\omega_1-\epsilon_{mn}+i\xi}
E_\alpha^{\omega_1} E_\beta^{\omega_2}
+
\sum_{\omega_1\omega_2}
\sigma^{(2;ee)}_{\lambda\alpha\beta}(m=n)
E_\alpha^{\omega_1} E_\beta^{\omega_2},
\end{align}
where we have used Eq.(\ref{sigmaee1}), which is derived in Appendix \ref{sigmaeemeqn}.
In addition, we have used the following relations
\begin{align}
\dfrac{\omega_\Sigma+2i\xi}{(\omega_1-\epsilon_{nm}+i\xi)(\omega_2-\epsilon_{mn}+i\xi)}
=
\dfrac{1}{\omega_2-\epsilon_{mn}+i\xi}
+
\dfrac{1}{\omega_1-\epsilon_{nm}+i\xi}
\\
\dfrac{\Delta^\lambda_{nm}}{(\omega_2-\epsilon_{nm}+i\xi)^2}
=
\partial_\lambda
\left(
\dfrac{1}{\omega_2-\epsilon_{nm}+i\xi}
\right),
\qquad
\partial_\lambda (r^\beta_{nm}r^\alpha_{mn})
=
\mathcal{D}^\lambda_{nm}(r^\beta_{nm})r^\alpha_{mn}
+
r^\beta_{nm}(\mathcal{D}^\lambda_{mn}r^\alpha_{mn})
\label{A3result}
\end{align}
to obtain the last line.

Finally, combining Eq.(\ref{Dresult}), Eq.(\ref{A1result}), Eq.(\ref{A2result}) with Eq.(\ref{A3result}),
we find
\begin{align}
J^{(2;0)}_\lambda = D+A_1+A_2+A_3
&=
\sum_{\omega_1\omega_2}
\left[
\sigma^{(2;ee)}_{\lambda\alpha\beta}(m=n)
+
\sigma^{(2;ee)}_{\lambda\alpha\beta}(m \neq n)
+
\sigma^{(2;ie)}_{\lambda\alpha\beta}
+
\sigma^{(ie;0)}_{\lambda\alpha\beta}
\right]
E_\alpha^{\omega_1} E_\beta^{\omega_2}
\nonumber \\
&=
\sum_{\omega_1\omega_2}
\left[
\sigma^{(2;ee)}_{\lambda\alpha\beta}
+
\sigma^{(2;ie)}_{\lambda\alpha\beta}
+
\sigma^{(ie;0)}_{\lambda\alpha\beta}
\right]
E_\alpha^{\omega_1} E_\beta^{\omega_2},
\end{align}
which gives the Eq.(\ref{intrinsicAC}) in the main text.

\section{The energy with the Luttinger-Kohn method}
\label{Kohnenergy}

The Hamiltonian $H=H_0+H_1$ satisfies the time-dependent Schr\"odinger equation 
\begin{align}
i\partial_t |\psi \rangle = H | \psi \rangle.
\label{SE1app}
\end{align}
Performing a unitary transformation for the quantum state: $|\psi\rangle \rightarrow e^{S}|\psi\rangle \equiv |\psi'\rangle$, we find:
\begin{align}
i\partial_t |\psi'\rangle 
=
(i\partial_t e^S)|\psi\rangle+e^{S} i \partial_t |\psi\rangle
=
(i\partial_t e^S)e^{-S}|\psi'\rangle+e^{S} H e^{-S} |\psi'\rangle
\equiv
H' | \psi'\rangle,
\end{align}
where we have used Eq. (\ref{SE1app}) and the Hamiltonian after this unitary transformation (denoted as $H'$) is explicitly given by:
\begin{align}
H \rightarrow H'\equiv
(i\partial_t e^S)e^{-S}+e^{S} H e^{-S}
=
e^S \left( -i\partial_t e^{-S} \right) + e^{S} H e^{-S}
\end{align}
where we have used that $0=\partial_t(e^{S}e^{-S})= \left( \partial_t e^{S} \right) e^{-S}+e^{S} \left( \partial_t e^{-S} \right)$.
Furthermore, using the operator identity:
\begin{align}
e^{A}Be^{-A}=B+\dfrac{1}{1!}[A,B]+\dfrac{1}{2!}[A,[A,B]]+\dfrac{1}{3!}[A,[A,[A,B]]] + \cdots,
\label{formal}
\end{align}
we find
\begin{align}
H'
&=
\left(1+S+\dfrac{1}{2}S^2 + \cdots \right) \left( - i\partial_t \right)
\left(1-S+\dfrac{1}{2}S^2 + \cdots \right)
+
\left( H+[S, H]+\dfrac{1}{2}[S,[S,H]] + \cdots \right)
\nonumber \\
&=
\left( i\partial_t S + S i\partial_t S - \dfrac{1}{2} i\partial_t S^2 \right)
+
\left( H+[S, H]+\dfrac{1}{2}[S,[S,H]] \right)
\nonumber \\
&=i\partial_t S + \dfrac{1}{2} [S, i\partial_t S] + H_0 + H_1 + [S, H_0] + [S, H_1]+\dfrac{1}{2}[S,[S,H]] 
\nonumber \\
&=
H_0+[S, H_1] + \dfrac{1}{2} [S, [S, H]+i\partial_t S],
\end{align}
where we only keep terms to the order of $S^2$ and choose
\begin{align}
i\partial_t S + H_1+ [S, H_0] = 0
\label{Smn0}
\end{align}
to fix the unitary transformation $S$.
We remark that this gauge choice bypasses the first-order energy correction in $H'$.
At this stage, by using the basis of $H_0|n\rangle=\epsilon_n|n\rangle$,
we further find that
\begin{align}
\langle n| \left( i\partial_t S + H_1+ [S, H_0] \right) |m \rangle = 0
\Longrightarrow
\partial_t S_{nm} + i\epsilon_{nm} S_{nm} = i E_\alpha (t) r^\alpha_{nm}, \quad (m \neq n)
\label{Smn}
\end{align}
where $E_\alpha (t) = \sum_{\omega_\alpha} E_\alpha e^{-i\omega_\alpha t} $ with $\omega_\alpha = \pm \omega$
and $r^\alpha_{nm}$ is the interband Berry connection. By solving Eq.(\ref{Smn}), we obtain:
\begin{align}
S_{nm} =  e^{-i\epsilon_{nm}t}
\left( \sum_{\omega_\alpha} i E_\alpha r^\alpha_{nm} \int e^{-i\omega_\alpha t} e^{\int i\epsilon_{nm} dt} dt \right)
=
-\sum_{\omega_\alpha} \dfrac{r^\alpha_{nm} E_\alpha e^{-i\omega_\alpha t} }{\omega_\alpha - \epsilon_{nm}}.
\label{Snm1}
\end{align}
As a result, the energy for $H'$ can be calculated as:
\begin{align}
E_n 
=
\langle n| H' | n \rangle 
&=
\epsilon_n + \langle n|[S, H_1] | n\rangle
+
\dfrac{1}{2} \langle n| [S, [S, H_0] + i\partial_t S]|n\rangle
\nonumber \\
&=
\epsilon_n + \langle n|[S, H_1] | n\rangle
-
\dfrac{1}{2} \langle n| [S, H_1] |n\rangle
\nonumber \\
&=
\epsilon_n
+
\dfrac{1}{2} \langle n| [S, H_1] |n\rangle
\nonumber \\
&=
\epsilon_n 
+
\dfrac{1}{2}
\sum_{\omega_\alpha}
\sum_{\omega_\beta}
\sum_{m}
\left(
-\dfrac{E_\alpha E_\beta r^\alpha_{nm} r^\beta_{mn} e^{-i(\omega_\alpha+\omega_\beta)t}}{\omega_\alpha-\epsilon_{nm}}
+
\dfrac{E_\alpha E_\beta r^\beta_{nm} r^\alpha_{mn} e^{-i(\omega_\alpha+\omega_\beta)t}}{\omega_\alpha-\epsilon_{mn}}
\right).
\end{align}
Note that we have used Eq.(\ref{Smn0}) and dropped $\dfrac{1}{2}\langle n|[S, [S, H_1]]|n\rangle \propto E_\alpha^3$.
Under the DC limit ($\omega_\alpha \rightarrow 0$), we obtain:
\begin{align}
E_n = \epsilon_n + \dfrac{1}{2} \sum_m \dfrac{g^{\alpha\beta}_{nm}E_\alpha E_\beta}{\epsilon_{nm}},
\end{align}
as given by Eq. (\ref{KLenergy}) in the main text.
Instead of directly taking the DC limit, if we first perform the frequency summation:
\begin{align}
E_n
=
\epsilon_n 
&+
\dfrac{1}{2}
\sum_{m}
\left(
-\dfrac{E_\alpha E_\beta r^\alpha_{nm} r^\beta_{mn} e^{-i2\omega t}}{\omega-\epsilon_{nm}}
+
\dfrac{E_\alpha E_\beta r^\beta_{nm} r^\alpha_{mn} e^{-i2\omega t}}{\omega-\epsilon_{mn}}
\right)
\nonumber \\
&+
\dfrac{1}{2}
\sum_{m}
\left(
-\dfrac{E_\alpha E_\beta r^\alpha_{nm} r^\beta_{mn}}{\omega-\epsilon_{nm}}
+
\dfrac{E_\alpha E_\beta r^\beta_{nm} r^\alpha_{mn}}{\omega-\epsilon_{mn}}
\right)
\nonumber \\
&+
\dfrac{1}{2}
\sum_{m}
\left(
-\dfrac{E_\alpha E_\beta r^\alpha_{nm} r^\beta_{mn}}{-\omega-\epsilon_{nm}}
+
\dfrac{E_\alpha E_\beta r^\beta_{nm} r^\alpha_{mn}}{-\omega-\epsilon_{mn}}
\right)
\nonumber \\
&+
\dfrac{1}{2}
\sum_{m}
\left(
-\dfrac{E_\alpha E_\beta r^\alpha_{nm} r^\beta_{mn} e^{+i2\omega t}}{-\omega-\epsilon_{nm}}
+
\dfrac{E_\alpha E_\beta r^\beta_{nm} r^\alpha_{mn} e^{+i2\omega t}}{-\omega-\epsilon_{mn}}
\right).
\end{align}
Under the DC limit, we obtain
\begin{align}
E_n = \epsilon_n + 2 \sum_m \dfrac{g^{\alpha\beta}_{nm}E_\alpha E_\beta}{\epsilon_{nm}},
\end{align}
as given in Ref. \cite{YBH2ndenergy}, where the additional factor $2$ arises from
the summation over two frequencies (Note that the electric field appears twice) but Ref. \cite{YBH2ndenergy} only considers one.
In the main text, the DC limit is directly taken without considering the frequency summation.

\twocolumngrid


\begin{thebibliography}{00}
\bibitem{Bfield}
Throughout this work, we will focus on the responses that is solely caused by the electric field $\vect{E}$,
but the semiclassical theory can also be applied to tackle with the responses
under the magnetic field $\vect{B}$.
\bibitem{vgauge}
We wish to mention that the perburbation Hamiltonian $H_1$
can also be treated by using the velocity gauge but both gauges are equivalent \cite{JEMoore2019}.
Throughout this work, the length gauge formulation of the response theory will be utilized.
\bibitem{footnote0}
We wish to mention that the last two terms in Eq.(\ref{Cn})
have been ignored in Ref. [\onlinecite{xiang}]
when formulate the third-order DC semiclassical theory with the wavepacket method.
\end{thebibliography}
\end{document}